\documentclass[aps,superscriptaddress,showpacs,twocolumn]{revtex4}
\usepackage{tabularx}
\usepackage{bm}
\usepackage{subfigure}
\usepackage{euscript}
\usepackage{epsfig,psfrag,subfigure}
\usepackage{graphicx,psfrag,subfigure}
\usepackage{color}
\usepackage{amsmath,amsfonts}
\usepackage{exscale}
\usepackage{amsbsy}
\usepackage{subfigure}
\usepackage{units}
\usepackage{epstopdf}

\begin{document}

\preprint{INT-PUB-09-024}
\title{Theory of combustion in disordered media}
\author{Mauro Schiulaz}
\author{Christopher R. Laumann}
\affiliation{Department of Physics, Boston University, Boston, MA, 02215, USA}
\author{Alexander V. Balatsky}
\affiliation{Institute for Materials Science, Los Alamos, NM 87545, USA}
\affiliation{Nordita, Center for Quantum Materials, KTH Royal Institute of Technology and Stockholm University, Roslagstullsbacken 23, SE-106 91 Stockholm, Sweden}
\author{Boris Z. Spivak}
\affiliation{Department of Physics, University of Washington, Seattle, WA 98195, USA}

\begin{abstract}
The conventional theory of combustion describes systems where all of the parameters are spatially homogeneous. 
On the other hand, combustion  in disordered explosives has long been known to occur after local regions of the material, called ``hot spots'', are ignited. 
In this article we show that a system of randomly distributed hot spots exhibits a dynamic phase transition, which, depending on parameters of the system  can be either first or second order. 
These two regimes are separated by a tri-critical point. 
We also show that on the qualitative level the phase diagram of the system is universal. 
It is valid in two and three dimensions, in the cases when the hot spots interact either by heat or sound waves and in a broad range of microscopic disorder models.
\end{abstract}

\date{\today}
\maketitle


\section{Introduction} 
\label{sec:introduction}

The microscopic mechanism of ignition of explosives is thermal in origin: it is related to exothermic reactions whose rates are quickly increasing functions of  temperature.
The theory of burning in uniform media was developed in Refs. ~\cite{Landau:1987ab,dremin1999toward,lee2008the,Frank-Khamenetski:1955aa,Zeldovich:1980aa}.
Usually two regimes are considered:
in the detonation regime heat propagates supersonically via shock waves,
while in the deflagration regime  (``combustion'') it propagates subsonically as determined by heat diffusion, and can be described by a nonlinear heat conduction equation.

The conventional theory of combustion describes systems where the parameters of the material are spatially homogeneous.
Generally however, in solid explosives these parameters are random, sample specific functions of coordinates.
Moreover, the explosion in disordered explosives has long been known to occur after local regions of the material, called ``hot spots'', are ignited by various processes
(see, e.g., Refs.~\cite{bowden1985initiation,Field:1992aa,Tarver:1996aa,McGrane:2009aa}).
In this case most of the predictions of the theory of burning are quite different from the conventional case \cite{Landau:1987ab,dremin1999toward,lee2008the,Frank-Khamenetski:1955aa}.
If the temperature near hot spot $i$ exceeds the critical value $T_{ci}$, a local explosion begins and the hot spot burns and releases energy $Q_{i}$.
This heat propagates through the neutral medium to other hot spots which it may in turn ignite.
If the hot spots are distributed sufficiently far from each other, the burning time of a hot spot is negligible compared to the time of intersite heat propagation and the dynamics are governed by the heat equation with pointlike sources:
\begin{align}
\label{eq:heat1}
(\partial_{t}-\frac{\kappa}{C} \partial_{{\bf r}}^{2})T({\bf r},t)=\sum_{i}\frac{Q_{i}}{C}\delta({\bf r}-{\bf r}_{i})\delta(t-t_{i})-\frac{T-T_{0}}{\tau}
\end{align}
where $t_i$ is the time of explosion of hot spot $i$. 

Generally, both $T_{ci}$ and $Q_{i}$ are random quantities.
We have shown in Ref.~\cite{MauroBorisChrisSasha} that in the deflagration regime a two-dimensional system of randomly distributed hot spots exhibits a 
dynamic phase transition. 
Once started, an explosion either is able to propagate through the entire sample or stops after burning only a finite fraction of the system.
Depending on the parameters of the system, the phase transition can be either a first-order transition or a second-order transition, which are separated by a tricritical point.
In the second-order regime, the transition lies in the percolative universality class.
The microscopic structure of hot spots is currently under debate (see, for example, a corresponding discussion in Ref.~\cite{bowden1985initiation}.) 
Different models yield different correlations between $T_{ci}$ and $Q_{i}$. The analysis in Ref.~\cite{MauroBorisChrisSasha} was performed for the case where the critical temperatures of the hot spots and the energies they release are proportional to each other $Q_{i}\sim T_{ci}$. 

In this article we argue that the results obtained in Ref.~\cite{MauroBorisChrisSasha} are in a sense universal.
They remain qualitatively the same for a much broader class of disorder models (as characterized by the correlations between $T_{ci}$ and $Q_{i}$).
They are also valid both in the three-dimensional (3D) case and in the case where a 2D layer of an explosive is embedded into a 3D neutral environment.
Perhaps most strikingly, they are valid even if the hot spots interact via  pressure waves rather than diffusive heat waves as we discuss next.

Under certain circumstances explosives can be ignited by the gentlest of mechanical blows.
This suggests that in these cases the ignition of the hot spots is controlled by  
local pressure rather than the local temperature (see, for example, Ref. \cite{bowden1985initiation}).  
That is, hot spot $i$ is ignited when the local pressure reaches its critical pressure, $P_{ci}$, which is randomly distributed.  
Currently, the microscopic mechanism of ignition of hot spots by pressure is under debate. 
One possible mechanism is that the explosive contains gas bubbles which can be  compressed.   
This compression leads to rapid adiabatic heating which in turn ignites the bubble.
In the regime of pressure-controlled ignition, the hot spots interact via sound or weak shock waves.
This regime may be viewed as intermediate between deflagration and detonation.
We show below that in this case, on the  qualitative level, the picture of explosion obtained in Ref.~\cite{MauroBorisChrisSasha} for the deflagration regime  remains the same: depending on the values of the parameters the system exhibits either a second- or first-order dynamical phase transition, separated by a tricritical point. 

The remainder of this paper is structured as follows. In Sec.~\ref{sec:deflagration} we introduce a model in which hot spots are activated by heat (deflagration) and discuss its phase diagram. In Sec.~\ref{sec:combustion_in_the_ballistic_regime} we do the same for a model in which hot spots are activated by pressure only. In Sec.~\ref{sec:numerical_methods} we describe in detail the numerical procedures we employ, and how the data are analyzed. Finally, in Sec.~\ref{sec:conclusion} we summarize our results and discuss some still open questions.


\section{Combustion in the deflagration regime}
\label{sec:deflagration}

In the deflagration regime, the energy released by the hot spots propagates via heat diffusion.
The analysis that follows is explained in more detailed in Ref.~\cite{MauroBorisChrisSasha}.
If the time it takes individual hot spots to burn is short compared to the time of the inter-hot spot heat propagation, then the combustion is described by the equation
\begin{align}
\label{eq:heat1}
(\partial_{t}-\frac{\kappa}{C} \partial_{{\bf r}}^{2})T({\bf r},t)=\sum_{i}\frac{Q_{i}}{C}\delta({\bf r}-{\bf r}_{i})\delta(t-t_{i})-\frac{T-T_{0}}{\tau}
\end{align}
Here $t_{i}$ is the time at which the temperature at the $i$th hot spot reaches its critical value $T({\bf r}={\bf r}_{i})=T_{ci}$, and ${\bf r}_{i}$ is the position of the spot. 
$C$ is the heat capacity per unit volume, and $\kappa$ is the heat conductance. $T_0$ is the temperature of the environment, while the dissipation time $\tau$ sets the strength of the coupling to the environment. 
For simplicity of notation, we set $T_0 = 0$ for the remainder of the paper.

The heat released from the explosion of a single hot spot at the origin of a uniform medium propagates as,
\begin{align}
	T(\mathbf{r}, t) &= \frac{Q/C}{(4\pi D t)^{d/2}} \exp\left(-\frac{r^2}{4Dt} - \frac{t}{\tau} \right),
\end{align}
where $d$ is the dimensionality of space,   $D = \kappa/C$ is the diffusion constant, and $r$ is the distance from the origin.
This wave ignites hot spots at position $\mathbf{r}_i$ if the local temperature rises above the critical temperature $T_{ci}$.
At a given position ${\bf r}$, the passing wave attains its maximum temperature
\begin{align}
	T_{max}(r,Q)&=&\frac{2^d Q}{\left(\sqrt{16 r^2 D \tau+d^2 D^2 \tau^2}-d D \tau\right)^{\frac{d}{2}}}\nonumber\\&\times&\exp\left(-\frac{\sqrt{16 r^2+d^2 D\tau}}{2\sqrt{D\tau}}\right).\label{eq:Tmax}	
\end{align}
at time
\begin{align}
	t^* = \frac{d \tau}{4}(\sqrt{\frac{4}{d^2} \frac{r^2}{l^2} + 1} - 1),
\end{align}
where $l = \sqrt{D\tau}$ is the dissipation length of the system.

To be concrete we use a distribution function of  $T_{ci}$ in the following form:
\begin{equation}
P\left(T_{ci}\right)\propto\begin{cases}
	T_{ci}^{\alpha}(T_{\textrm{max}}-T_{ci})^{\alpha} & 0\leq T_{ci} \leq T_{\textrm{max}},\\
	0 & \textrm{otherwise},
\end{cases}
\label{eq:PTc}
\end{equation}
where we normalize the units of temperature such that $T_{\textrm{max}} = 1$ and have chosen $\alpha = 4$ for all simulations, to provide the soft gap necessary for stability of the sample.

Generally, the values of $T_{ci}$ and $Q_{i}$ are correlated.  To illustrate the universality of the phase diagram we consider below three models for the correlation, wherein $Q_{i}$ is either linearly, quadratically, or inversely proportional to $T_{ci}$:
 $Q_{i}=BCa^3T_{ci}$,
$Q_{i}=Ba^3CT_{ci}^{2}/T_\textrm{max}$, or $Q_{i}=BT_\textrm{max}^2a^3C(T_{ci}+U)^{-1}$. 
Here $C$ is the heat capacity, $a$ is the average distance between  the hot spots, $U$ is a cutoff temperature, and $B$ is a dimensionless constant characterizing the strength of the hot spot's explosions. 
 
 The relation between the dissipation length $l$  the average distance between the hot spots $a$, and the parameter $B$ determines the dynamics of the burning.
If $l\ll a$, then each hot spot can receive heat only from its closest neighbors and the effects of heat accumulation from multiple explosions may be neglected.
It has been shown in Ref. ~\cite{MauroBorisChrisSasha} that in this limit, the theory of the heat propagation in the system of hot spots can be reduced to   percolation theory.
Thus  for $B>B_{c}$, the explosion percolates through the sample, while for $B<B_c$  the explosion propagates up to the correlation radius
\begin{equation}\label{corrad}
	 \xi\sim |B-B_{c}|^{-\nu}.
\end{equation}
In this regime, the explosion activates only a finite number of hot spots, and
\begin{equation}
 \bar{N}_{\textrm{exp}}\sim \frac{1}{|B-B_c|^{\gamma}}
\end{equation}
increases as a power law of $B-B_{c}$.  
The value of the exponents $\nu$ and $\gamma$ are given by percolation theory.
 
At $B>B_{c}$ the explosion propagates along the percolating cluster and burns a finite fraction of the system. We call this the explosive phase.
As a result the system exhibits a second order phase transition.

This picture changes for $l\gg a$: in this regime heat from multiple sources accumulates, and no mapping to percolation theory is possible: rather, the system displays a first-order transition, in which the explosion travels either to distances of order $a$ only or through the entire sample. No diverging length scale can be identified at the transition point. The two regimes are separated by a tricritical point.

\begin{widetext}

\begin{figure}
\begin{minipage}[t]{0.4\linewidth}
\includegraphics[scale=0.7]{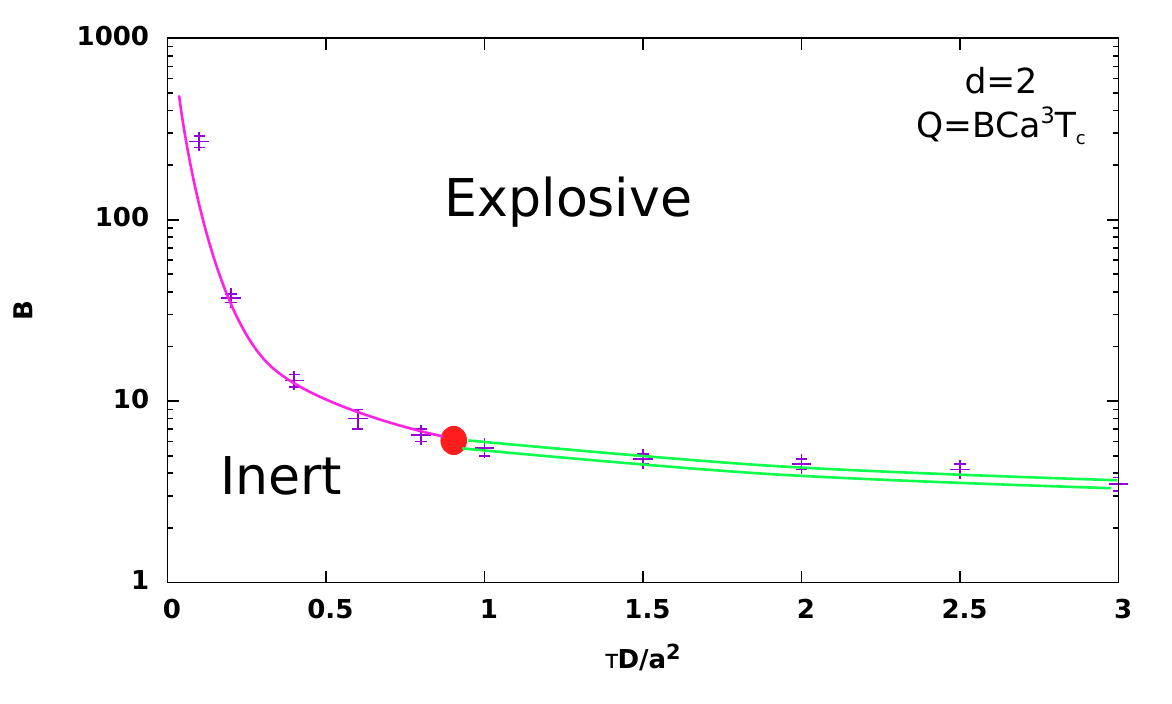}
\includegraphics[scale=0.7]{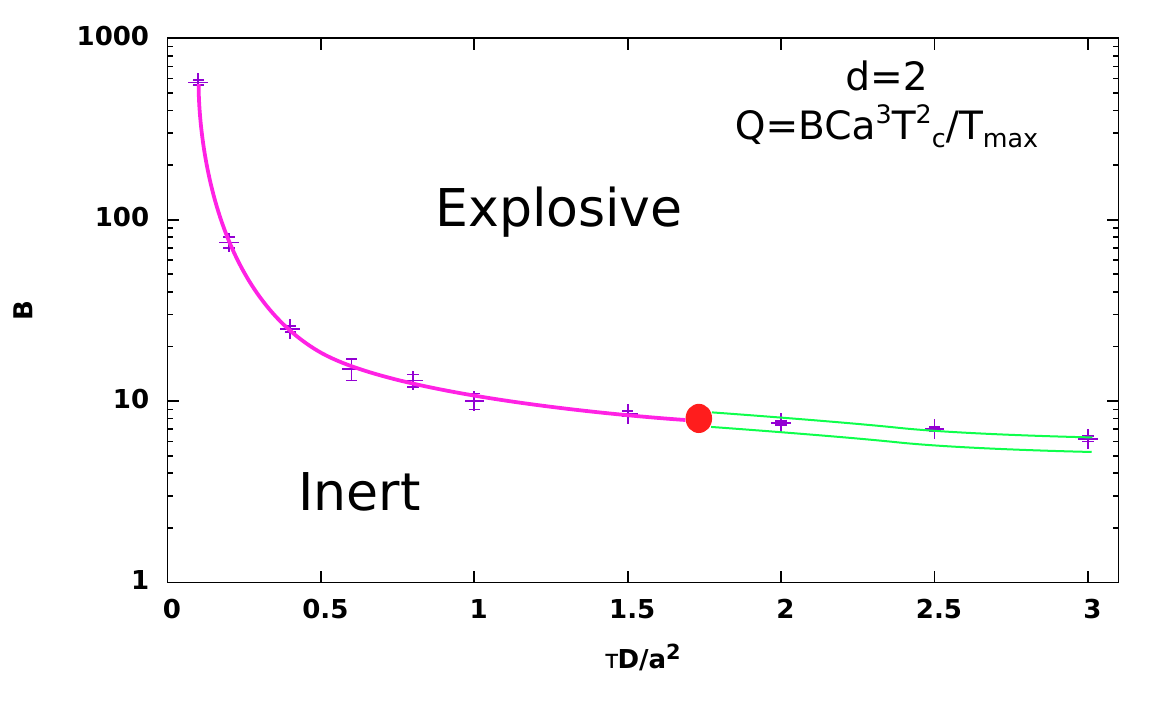}
\includegraphics[scale=0.7]{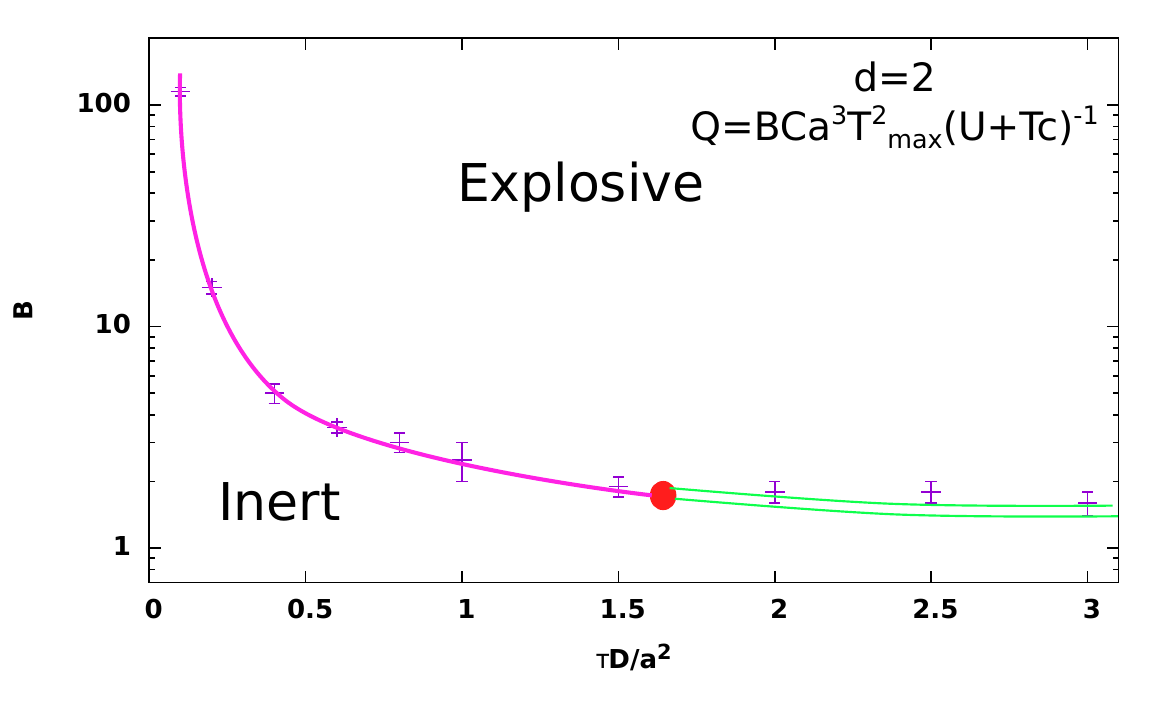}
\end{minipage}
\hspace{1.5cm}
\begin{minipage}[t]{0.4\linewidth}
\includegraphics[scale=0.7]{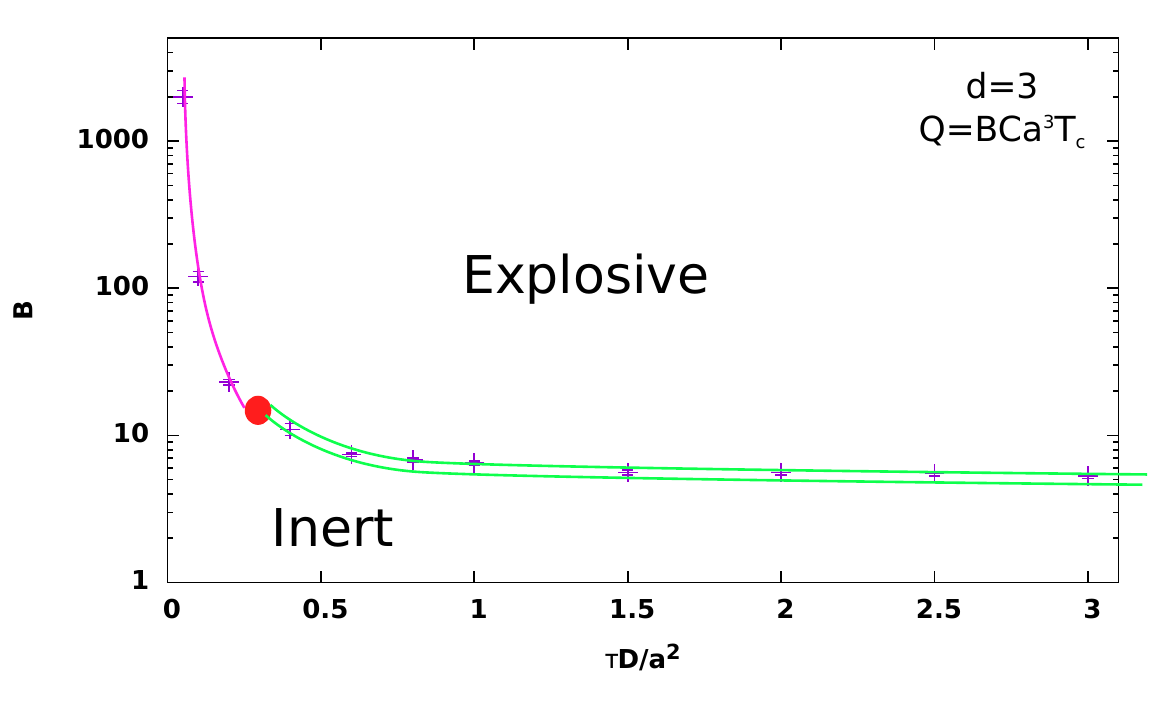}
\includegraphics[scale=0.7]{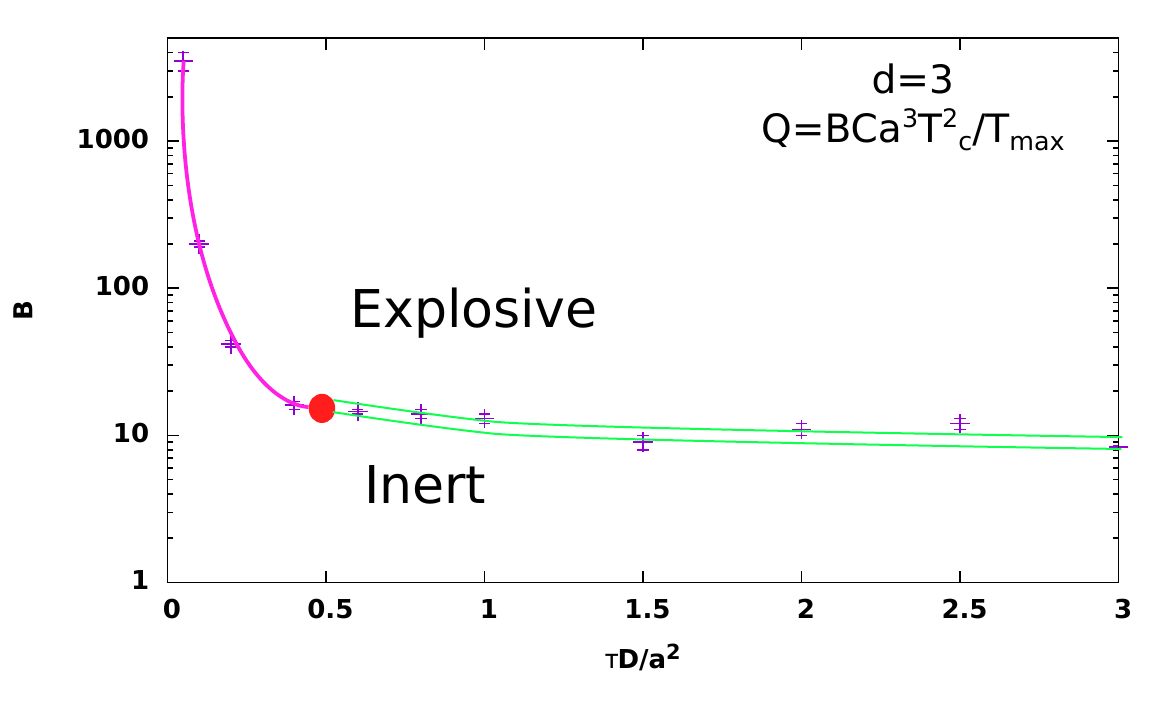}
\includegraphics[scale=0.7]{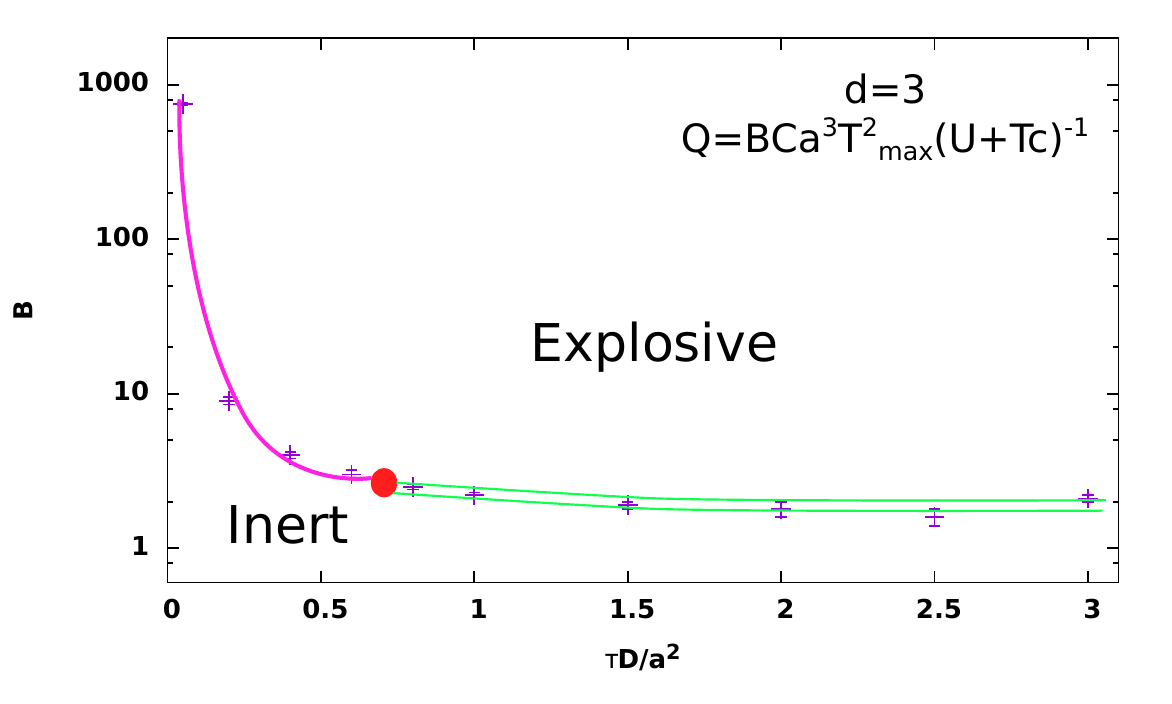}
\end{minipage}
\caption{Phase diagram for the deflagration model, in spatial dimension $d=2$ (a-c) and $d=3$ (d-f). The models with $Q\propto T_c$ (a,d), $Q\propto T_c^2$ (b,e), and $Q\propto (T_c+U)^{-1}$ (c,f) are shown. A tricritical point (red dot) separates a second-order transition line (purple single line) from a first-order one (green double line).}
\label{fig:deflagration_phasediagram}
\end{figure}

\end{widetext}

In Fig.~\ref{fig:deflagration_phasediagram} we present phase diagrams of the system obtained by numerical simulations of the three different correlation models for the $T_{ci}$ and $Q_i$.  
Thus we argue that on the qualitative level, this diagram remains the same for a broad range of correlations and in different dimensionalities.
 
One may also wonder whether the results would change if one considers the more realistic situation of a two-dimensional array of hot spots embedded in a three-dimensional environment. 
In this case, the surrounding environment absorbs heat from the system, thereby providing a source of dissipation even when $\tau=\infty$. 
This is not enough to make the transition become second-order, though: as we show in Sec.~\ref{sec:numerical_methods}, in the absence of dissipation the transition is first-order even in this case (albeit the discontinuity is weaker than in the purely two dimensional model). Therefore, one needs to actively drain heat from the system, in order to observe the second-order transition.

\section{Combustion in the pressure-mediated regime} 
\label{sec:combustion_in_the_ballistic_regime}

In this section we consider the case where the explosion of a hot spot creates a spherical sound wave which propagates through a neutral media igniting other hot spots. 
In the absence of the dissipation, the system of hot spots can be described by a sound wave equation for the pressure $P({\bf r},t)$:
\begin{equation}\label{eq:wavepropagation}
(\partial^{2}_t -c^2\partial^{2}_{{\bf r}})P({\bf r},t)= \sum_{i}Q_{i}e^{-\frac{({\bf r}-{\bf r}_{i})^2}{r_{0}^{2}}}\delta(t-t_{i}).
\end{equation}
As in the deflagration regime, $t_i$ is the time at which the pressure $P({\bf r}_i,t)$ is larger than the critical pressure $P_{ci}$, while $c$ is the speed of sound in the medium and $Q_i$ is the magnitude of the pressure impulse (rather than a heat released). 

Even at late times, sound waves are more sensitive to the initial profile of an impulse than heat waves are. 
At distances larger than the spot radius $r_{0}$ in $3$D, a wave created by an explosion of a hot spot has the form of a spherical shell with width of order $r_{0}$.
In Eq.~\eqref{eq:wavepropagation}, we take into account this spatial structure by using a Gaussian pressure impulse of width $r_0$  rather than a delta function.
This has the added advantage that it reduces simulation errors due to the discretization of space (which modifies the high-momentum dispersion of the waves). 

Equation~(\ref{eq:wavepropagation}) describes spherical sound waves propagating from some localized sources. 
Such waves are well known to have a different character in two and three dimensions \cite{Landau:1987ab}. 
In $d=3$, assuming spherical symmetry around a hot spot located at $r=0$, we can rewrite Eq.~(\ref{eq:wavepropagation}) as
\begin{align}
\frac{\partial^2 P}{\partial t^2}(r,t)=c^2\frac{1}{r^2}\frac{\partial}{\partial r}\left(r^2\frac{\partial P}{\partial r}\right).
\end{align}
This equation can be solved by defining $\phi(r,t)\equiv rP(r,t)$, 
with
\begin{align}
\label{eq:solutionspherical}
P(r,t)=\frac{1}{r}\phi(r-ct),
\end{align}
where $\phi$ is a function, determined by  the initial conditions. For the Gaussian source term in Eq.~(\ref{eq:wavepropagation}), we have $\phi(r-ct)=\frac{Qr_0^2}{2c} \exp\left(-\frac{(r-ct)^2}{r_0^2}\right)$. 
Such a solution describes a way propagating outwards from the center, without changing its qualitative shape: in particular, only a Gaussian small residual pressure is left behind the wave front.
No such procedure is possible in two dimensions. Rather, it can be shown~\cite{Landau:1987ab} that, in this case, an outgoing solution of the wave equation with radial symmetry can be written as
\begin{align}
P(r,t)=\int_{-\infty}^{ct-r}d\xi\frac{\phi(\xi)}{\sqrt{(ct-\xi)^2-r^2}},
\end{align}
where again the function $\phi$ is determined by the initial conditions. In this case, the wave has no well defined backward front: it can be checked that the pressure decays in time as
\begin{align}
P\sim\frac{1}{t}
\end{align}
for $r\ll ct$, i.e., far behind the wave front.

Despite this difference, we find numerically that the phase diagram for the explosion model with sound waves is qualitatively similar to the one with deflagration. We use a distribution function of  $P_{ci}$ analogous to the one used in the deflagration regime,
\begin{equation}
P\left(P_{ci}\right)\propto\begin{cases}
	P_{ci}^{\alpha}(P_{\textrm{max}}-P_{ci})^{\alpha} & 0\leq P_{ci} \leq P_{\textrm{max}},\\
	0 & \textrm{otherwise},
\end{cases}
\label{eq:Pc}
\end{equation}
 where once again $P_{\textrm{max}} = 1$ and $\alpha = 4$.

 In Fig.~\ref{fig:Soundwaves} we present results of numerical simulations of Eq.~(\ref{eq:wavepropagation})  in both $d=2$ and $d=3$, with relations $Q_{i}=BcP_{ci}/a$, $Q_{i}= Bc^2 P_{ci}^{2}/(a^2 P_\textrm{max})$, and $Q_{i}=BcP_\textrm{max}/a (P_{ci}+U)^{-1}$, with the cutoff pressure $U=P_\textrm{max}/2$.  
 In all these cases, as a function of the coefficient $B$ the system of the hot spots exhibits a first order phase transition.  

Equation~(\ref{eq:wavepropagation}) does not take into account sound wave dissipation which is controlled by heat conduction. 
Since the propagating spherical waves have a single spatial scale (the width) of order $r_0$, the decay length of each wave is of order $l\sim r_0^{-2}$. 
Therefore, in this limit Eq.~\eqref{eq:solutionspherical} should be multiplied by a factor $\exp(-r/l)$.
If $l\ll a$, then a hot spot can be ignited only by waves generated by neighbors.  In this case, similarly to the case of the deflagration regime,  the problem can be reduced to a percolation problem, and the system exhibits a second-order dynamical phase transition. 
The only difference from the deflagration regime is that we have to introduce the critical pressures igniting the hot spots $P_{ci}$ instead of critical temperatures. 
Thus, on the qualitative level, the phase diagram of the system is similar to that presented in Fig.~\ref{fig:deflagration_phasediagram} for the deflagration regime.
This means that the structure of the phase diagram does not depend on the details of how energy is transported within the system, but rather is of universal character.

Also in this case, we have explored the case of a two-dimensional explosive embedded in a three-dimensional environment, finding that also in this case the transition is first order. This further confirms the universal character of our results.

\section{Numerical methods} 
\label{sec:numerical_methods}

In order to simulate disordered combustion, we follow and generalize the procedure described in Ref.~\cite{MauroBorisChrisSasha}. We generate samples composed of a two- or three-dimensional cubic lattice of pointlike hot spots embedded in a passive medium with uniform diffusivity. 
The lattice spacing is $a$ and we take periodic boundary conditions for all simulations, except for that of the 2D active system embedded in a 3D material, where we take periodic boundary conditions in the two lattice directions and absorbing boundary conditions in the transverse direction.

%

The heat released on exploding hot spot $i$, $Q_i$, is determined by $T_c^{(i)}$ according to which of the three microscopic disorder models we are simulating:
$Q_i = B T_{ci}$,  $Q_i = B T_{ci}^{2}$, or $Q_i = B (T_{ci} + U)^{-1} $. 
The units are such that $C = a = 1$. 
We have chosen the value $U=T_{\textrm{max}}/2$, to cut off the unphysical divergence of $Q$ for $T_c\rightarrow 0$ in model 3. 

For the pressure-mediated model, the samples are generated according to the above description simply replacing the critical temperatures, $T_{ci}$, by critical pressures, $P_{ci}$. 

\subsection{Numerical simulation of the deflagration model}
\label{sec:2ddeflagration}

For the diffusive model, we simulate the heat equation with relaxation, Eq.~\eqref{eq:heat1}, discretizing time in steps $\Delta t$.  We take a regular lattice of hot spots, with lattice spacing $a$. 
For each run, we initialize the combustion by making a single randomly chosen site explode, releasing its energy $Q_{i}$. 
During each time step, the heat propagates according to Eq.~(\ref{eq:heat1});
additionally, when the temperature at an unexploded site $i$ exceeds its local critical temperature, i.e., if $T({\bf r}_i,t)\geq T_{ci}$, site $i$ explodes and releases energy $Q_{i}$.
After exploding, the hot spot at site $i$ is exhausted and will not explode again.
Explicitly, this is implemented by iterating the following procedure at each time step $\Delta t$, for each site $i$.

\begin{itemize}

\item If site $i$ is still active and $\ensuremath{T({\bf r}_{i},t)\geq T_{ci}}$, the site explodes, 
\begin{equation}
T\left(\mathbf{r}_{i},t\right)\rightarrow T\left(\mathbf{r}_{i},t\right)+Q_{i},
\end{equation}
and becomes exhausted. 

\item The temperature then relaxes diffusively,
\begin{equation}
T\left(\mathbf{r}_{i},t+\Delta t\right)=\frac{T\left(\mathbf{r}_{i},t\right)+\sum_{j=1}^{Z}T\left(\mathbf{r}_{j},t\right)}{Z+1},
\end{equation}
where $Z$ is the coordination number of the lattice, and $j$ labels the nearest neighbors of site $i$. This update rule fixes the value of $\Delta t$:
\begin{align}
\Delta t=(Z+1)^{\frac{2}{d}}\frac{a^2}{4\pi D}.
\end{align}
The procedure is asymptotically correct although at short times it distorts the dynamics. 
However, it allows us to greatly increase the efficiency of the simulations, and we do not expect it to have any significant qualitative impact on the physics of the system. 
Its primary effect is to introduce a systematic deviation in the {\em position} of the transition point, but not to change its nature.

\item Finally, the heat dissipation acts:
\begin{equation}
T\left(\mathbf{r}_{i},t+\Delta t\right)\rightarrow T_0+(T\left(\mathbf{r}_{i},t+\Delta t\right)-T_{0})e^{-\frac{\Delta t}{\tau}}.
\end{equation}

\end{itemize}

We end the simulation when the temperature field has decayed to the point that no further sites may explode.
Our primary measurement is then to count the number $N_{\textrm{exp}}$ of sites which have exploded.

\begin{figure}[ptb]
\includegraphics[scale=0.7]{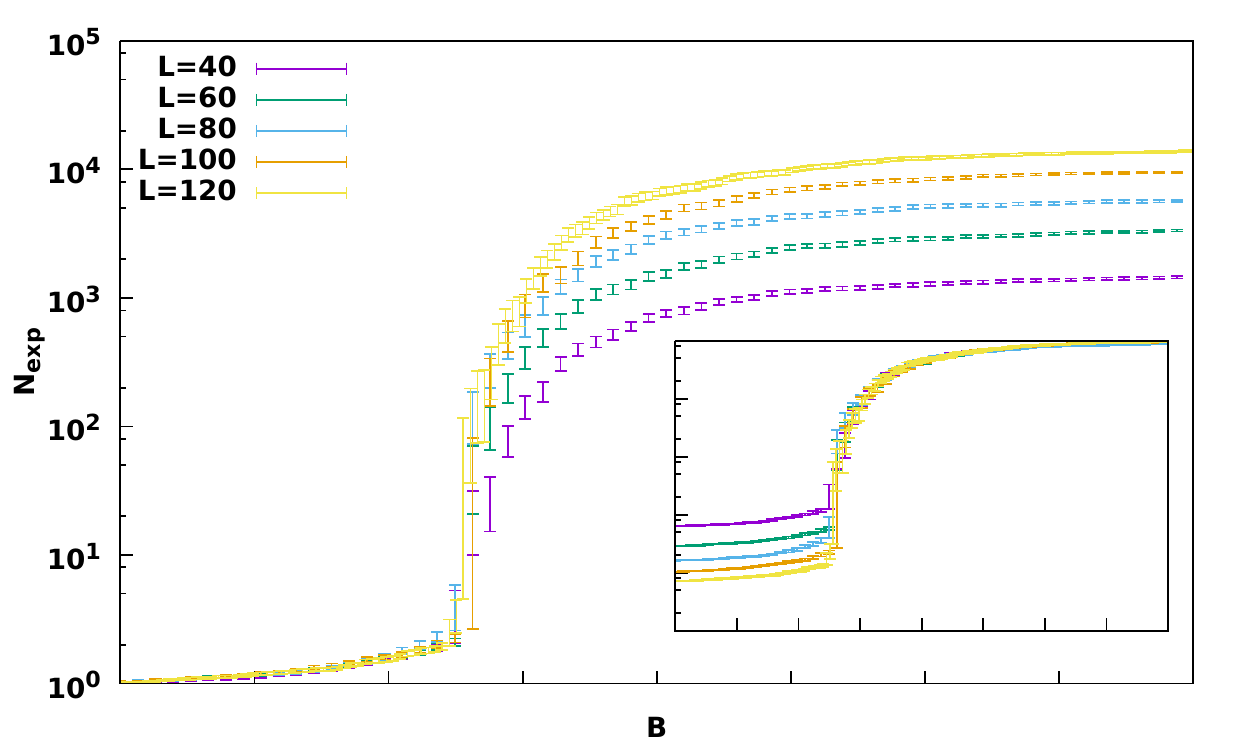}
\includegraphics[scale=0.7]{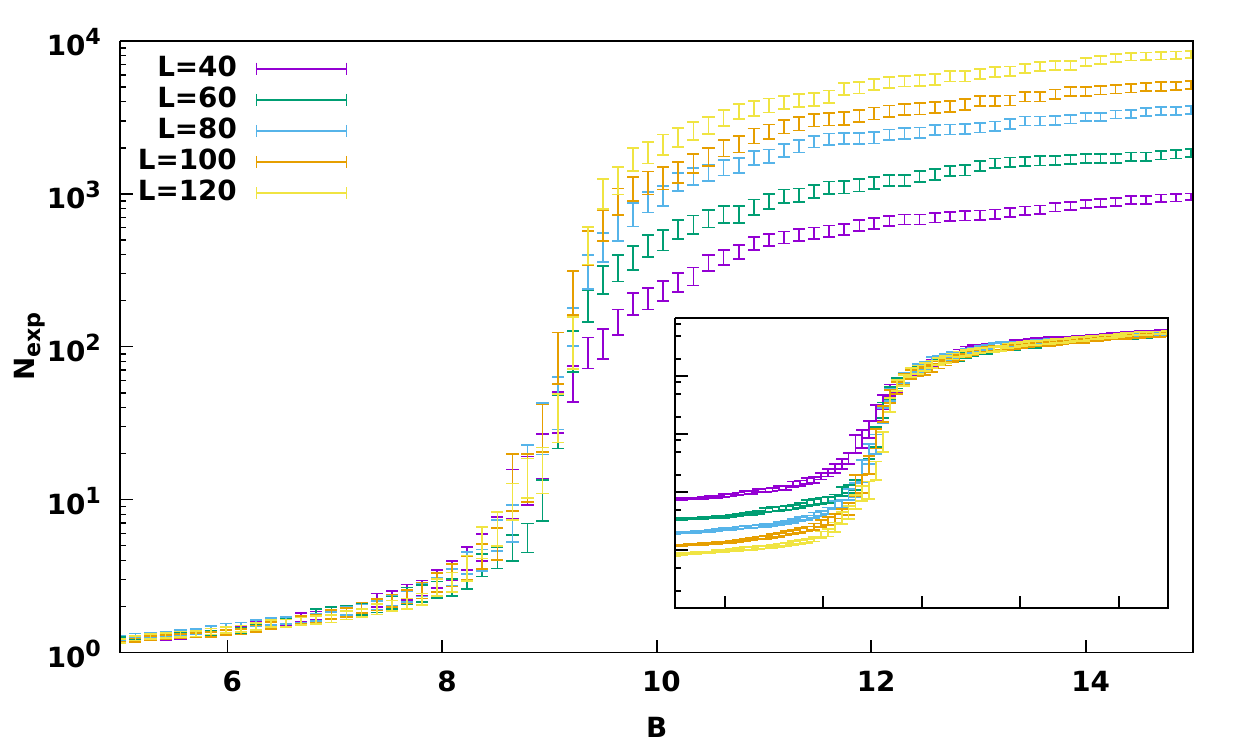}
\includegraphics[scale=0.7]{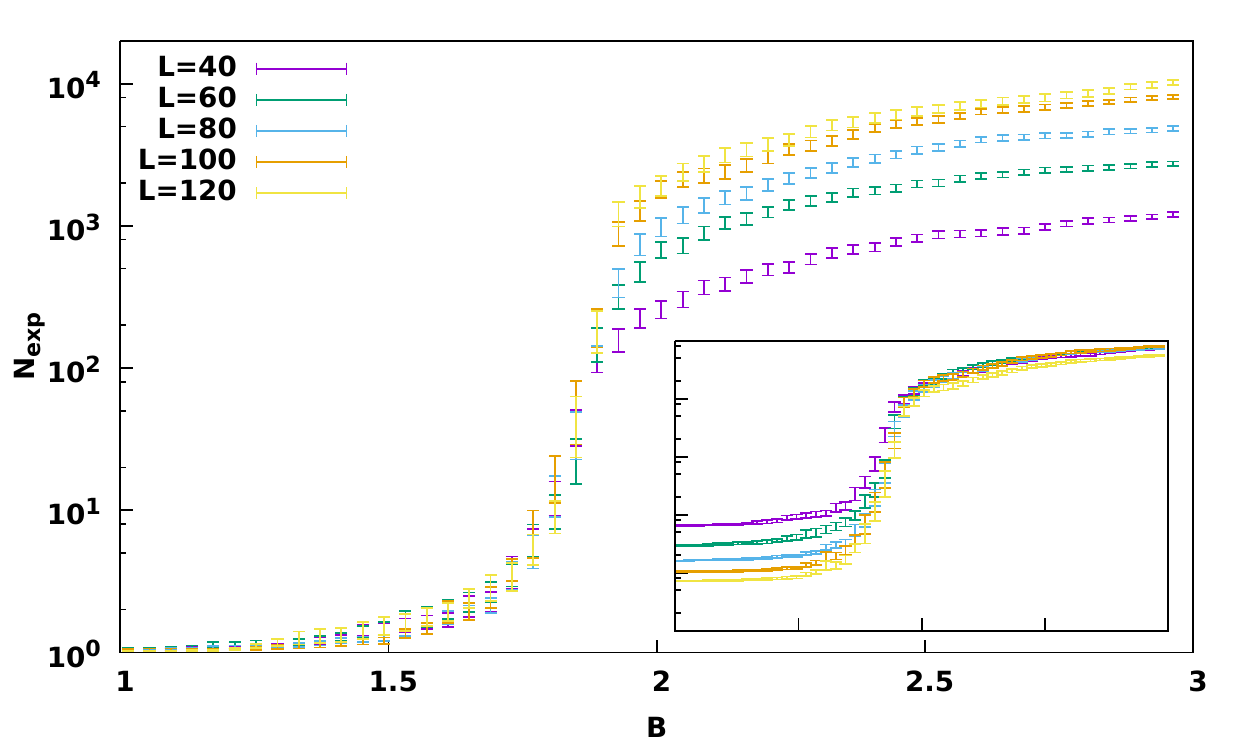}
\caption{
Deflagration regime, 2D explosive layer embedded in a 3D material. 
{\em Main plots:} Number of exploded sites $N_{\textrm{exp}}$ as a function of $B$, for dissipation time $\tau=\infty$. The models with $Q\propto T_c$ (a), $Q\propto T_c^2$ (b), and $Q\propto (T_c+U)^{-1}$ (c) are shown.  
{\em Insets:} A first-order transition is clearly visible and the quantity $N_{\textrm{exp}}/L^2$, plotted as a function of $B$, is system size independent after the transition.
}
\label{fig:Heat2Din3D}
\end{figure}

The numerical data are shown in Fig.~\ref{fig:Deflagration_3D} for the three-dimensional deflagration model, in Fig.~\ref{fig:Heat_2D} for the two-dimensional model, and in Fig.~\ref{fig:Heat2Din3D} for the two dimensional layer imbedded in a three-dimensional medium.  Since $N_{\textrm{exp}}$ remains finite in the inert phase, and becomes extensive in the exploded phase, we take the fraction of exploded sites $N_{\textrm{exp}}/L^d$ as an order parameter for the transition.

\begin{widetext}

\begin{figure}
\begin{minipage}[t]{0.4\linewidth}
\includegraphics[scale=0.7]{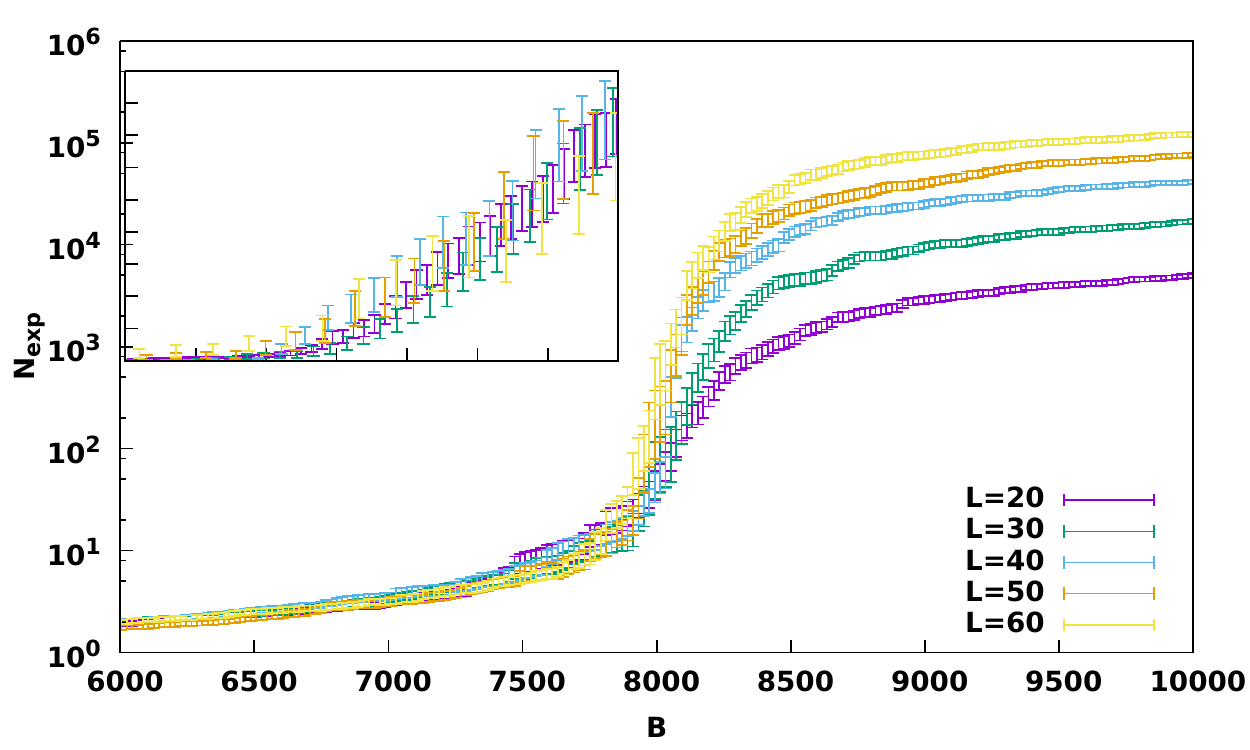}
\includegraphics[scale=0.7]{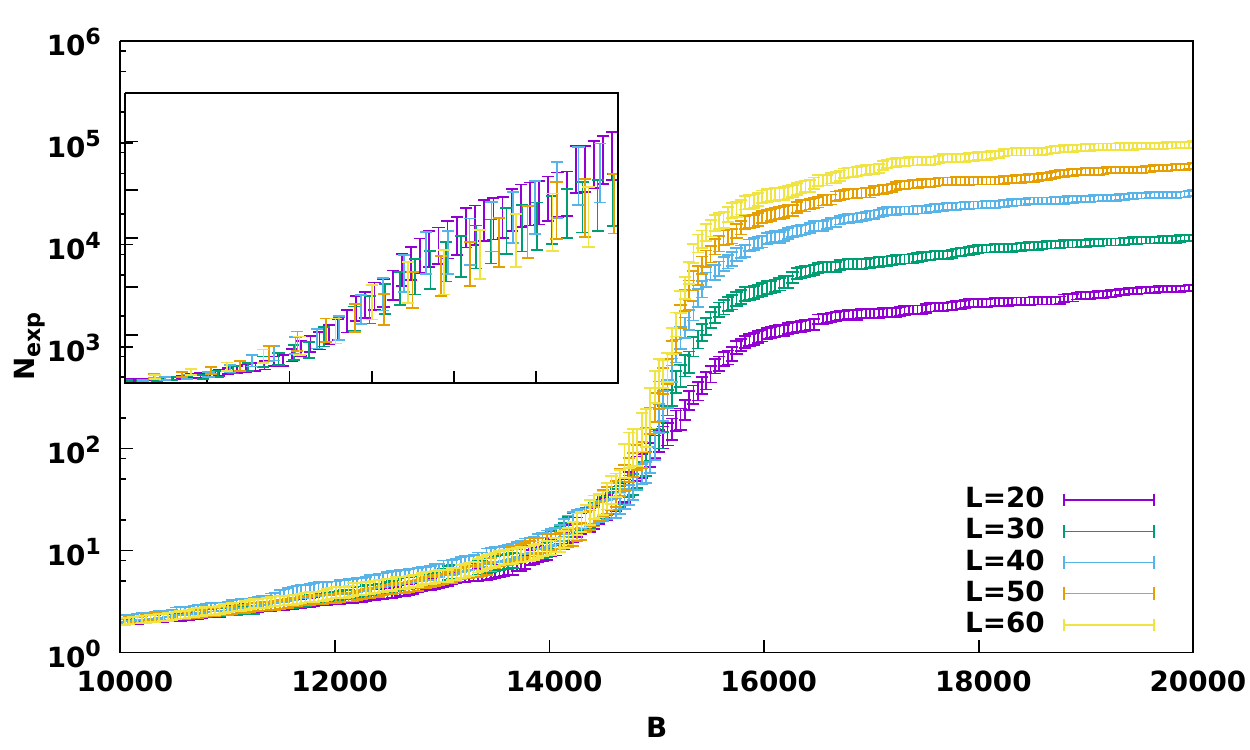}
\includegraphics[scale=0.7]{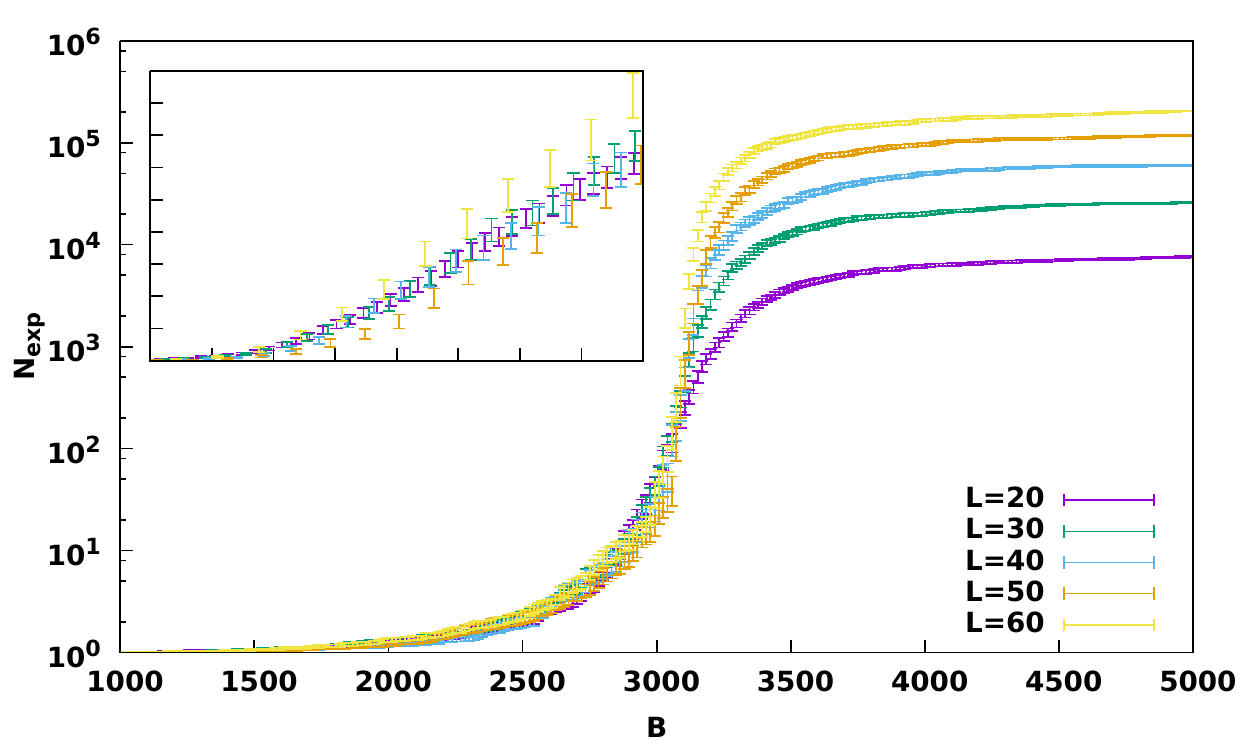}
\end{minipage}
\hspace{1.5cm}
\begin{minipage}[t]{0.4\linewidth}
\includegraphics[scale=0.7]{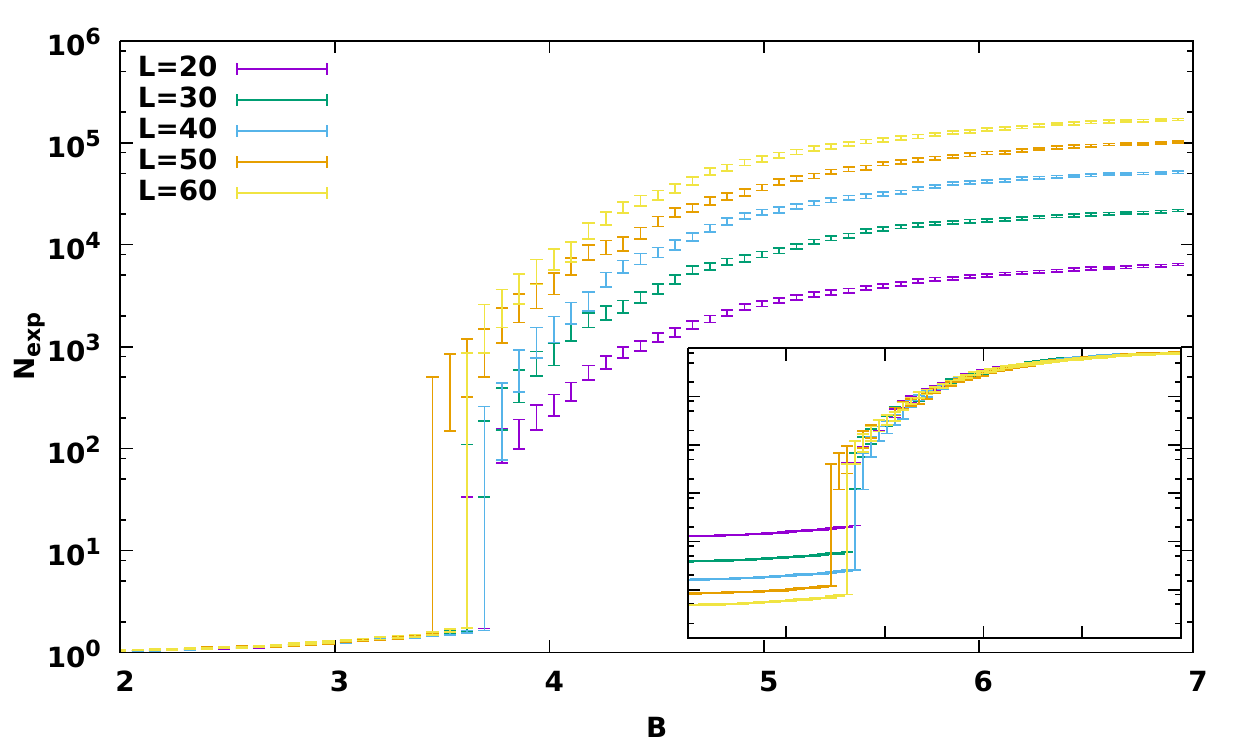}
\includegraphics[scale=0.7]{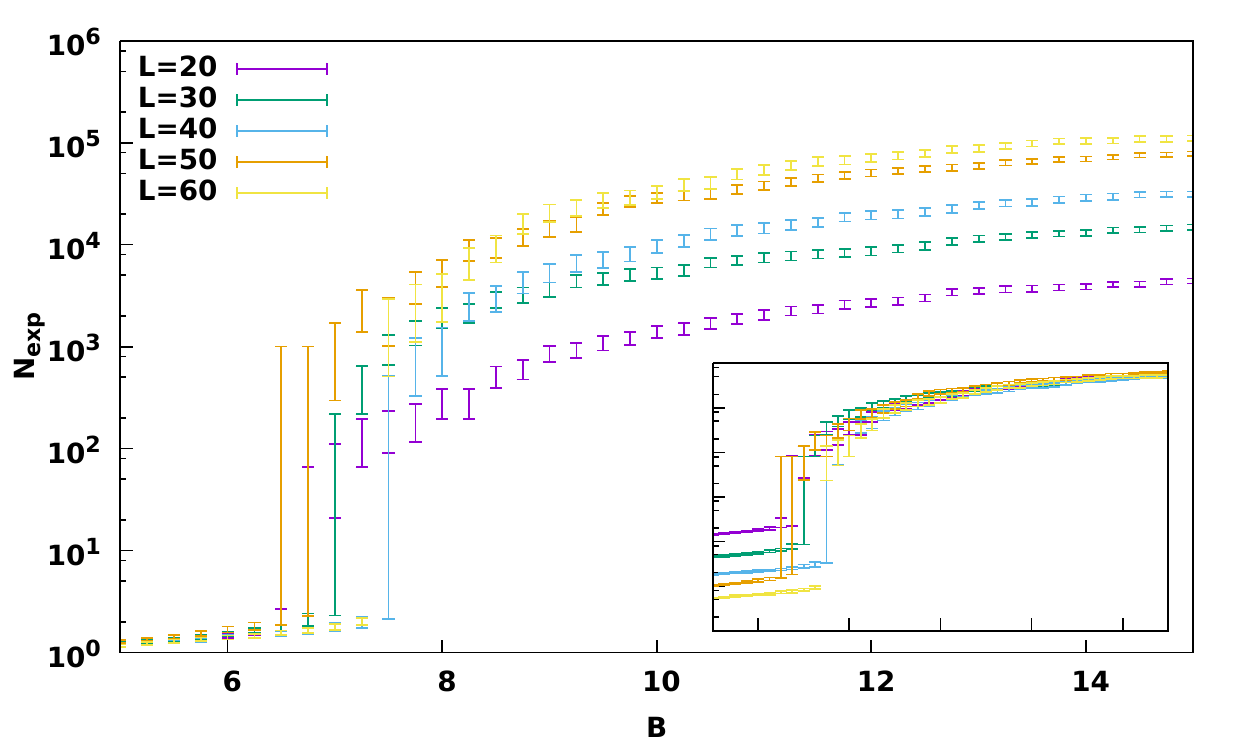}
\includegraphics[scale=0.7]{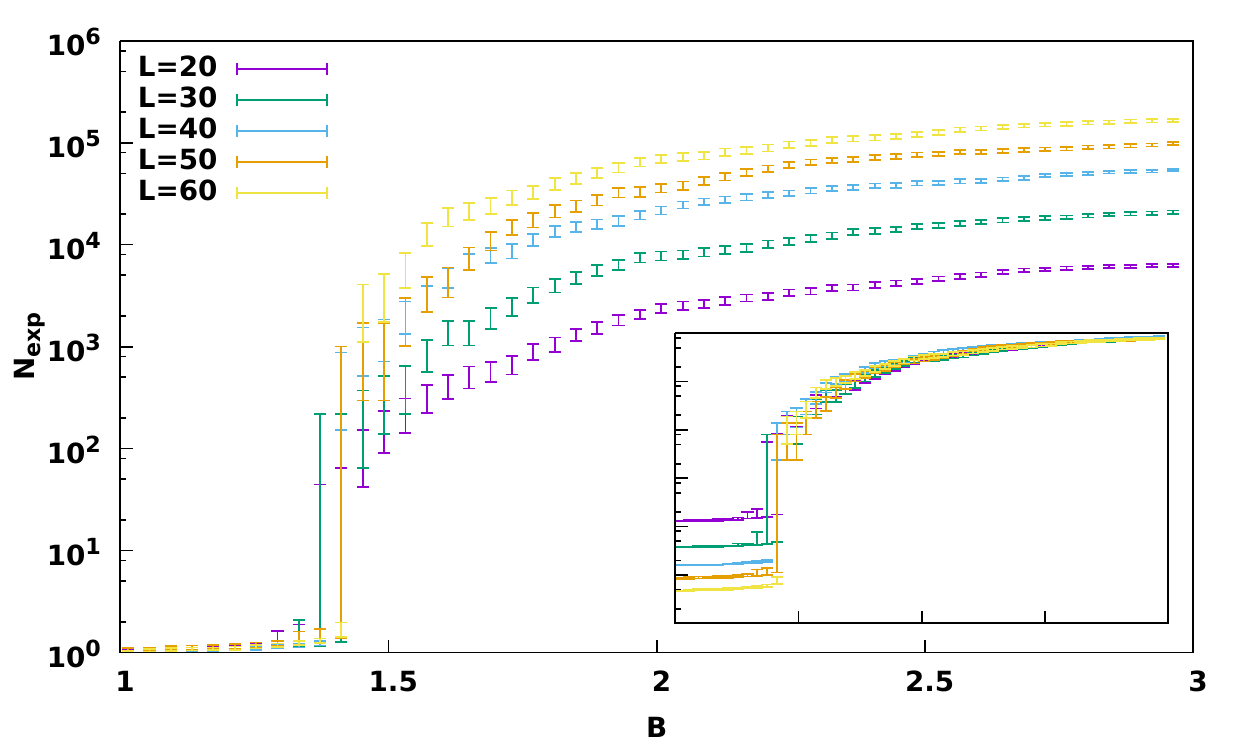}
\end{minipage}
\caption{Three-dimensional deflagration regime. {\em Main plots:} Number of exploded sites $N_{\textrm{exp}}$ as a function of $B$, for dissipation times $\tau=0.04 D/a^2$ (a-c) and $\tau=1.25 D/a^2$ (d-f). From top to bottom, the models with $Q_{i}\propto T_{ci}$ (a,d), $Q_{i}\propto T_{ci}^{2}$ (b,e), and $Q_{i}\propto (T_{ci}+U)^{-1}$ (c,f) are shown. Different data sets correspond to different system sizes $L$. {\em Insets:} for short dissipation times (a-c), finite size scaling shows that the transition falls into the universality class of 3D percolation. This is shown by plotting $N_\textrm{exp}L^{-\frac{\gamma}{\nu}}$ as a function of $L^{\frac{1}{\nu}}(B-B_c)/B_c$. For long dissipation times (d-f),  a first-order transition is clearly visible, and the quantity $N_{\textrm{exp}}/L^3$, plotted as a function of $B$, is system size independent after the transition.}
\label{fig:Deflagration_3D}
\end{figure}

\begin{figure}
\begin{minipage}[t]{0.4\linewidth}
\includegraphics[scale=0.7]{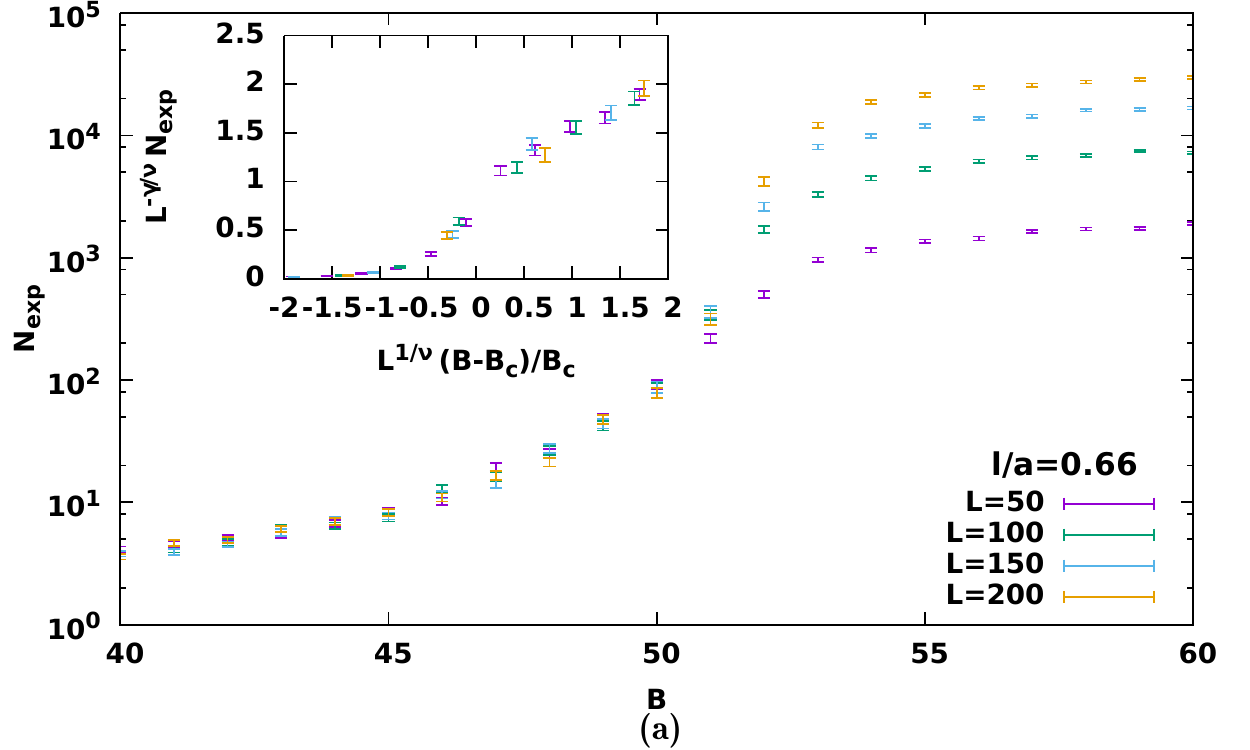}
\includegraphics[scale=0.7]{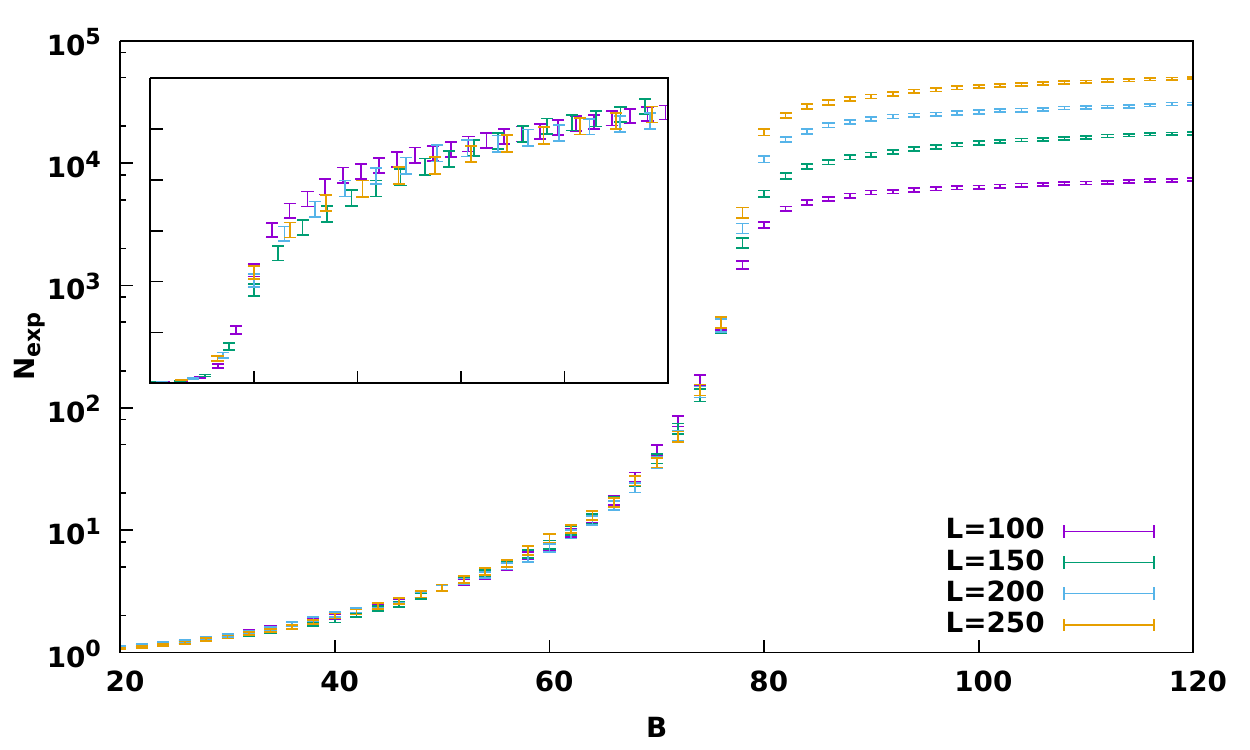}
\includegraphics[scale=0.7]{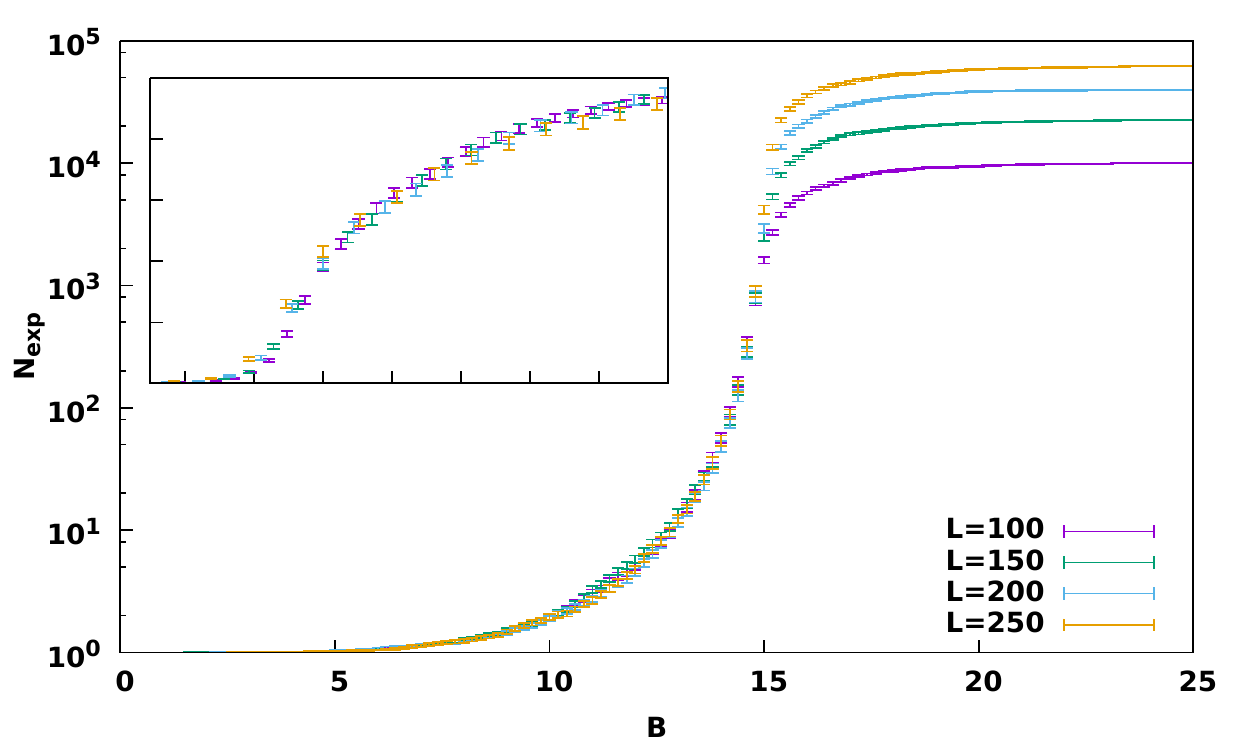}
\end{minipage}
\hspace{1.5cm}
\begin{minipage}[t]{0.4\linewidth}
\includegraphics[scale=0.7]{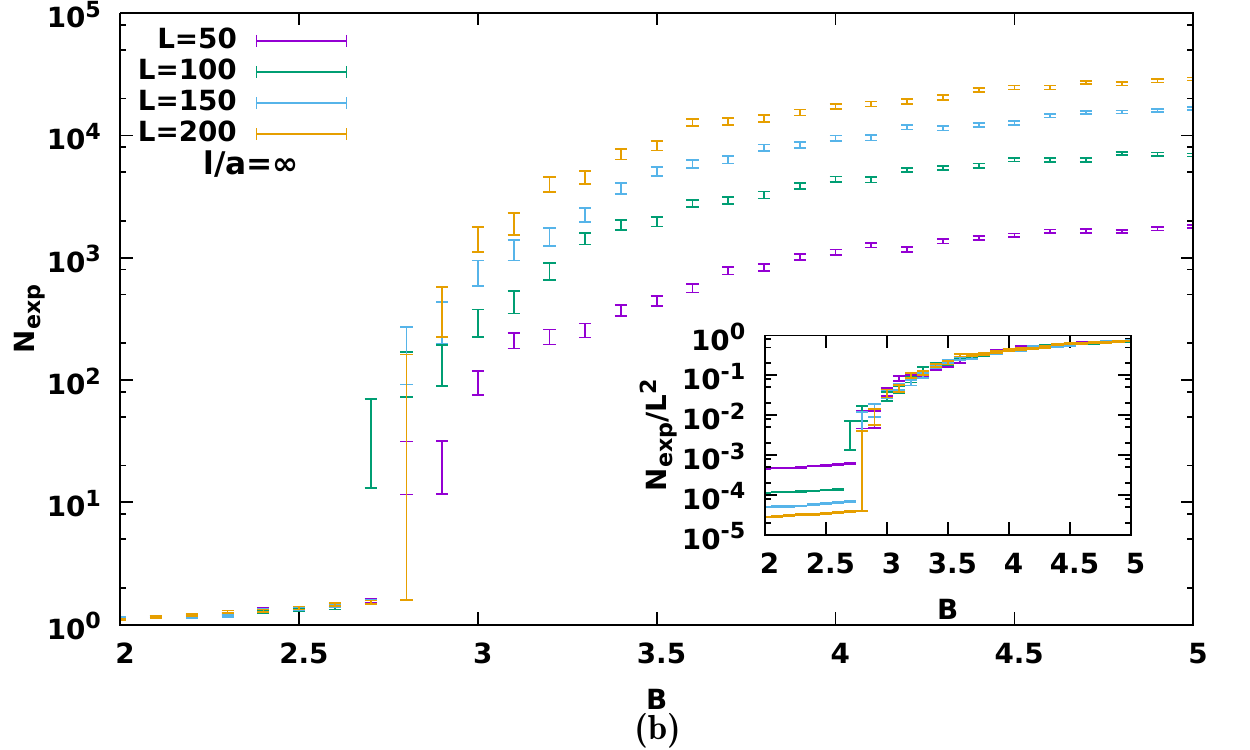}
\includegraphics[scale=0.7]{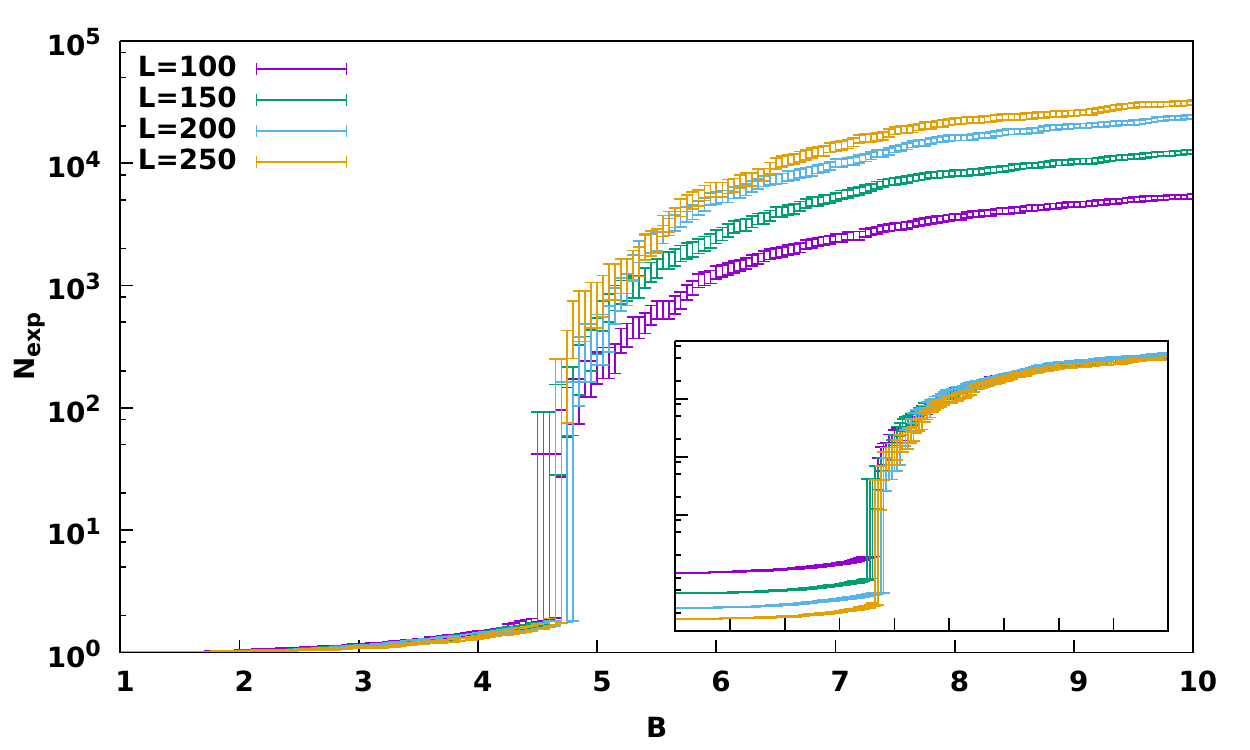}
\includegraphics[scale=0.7]{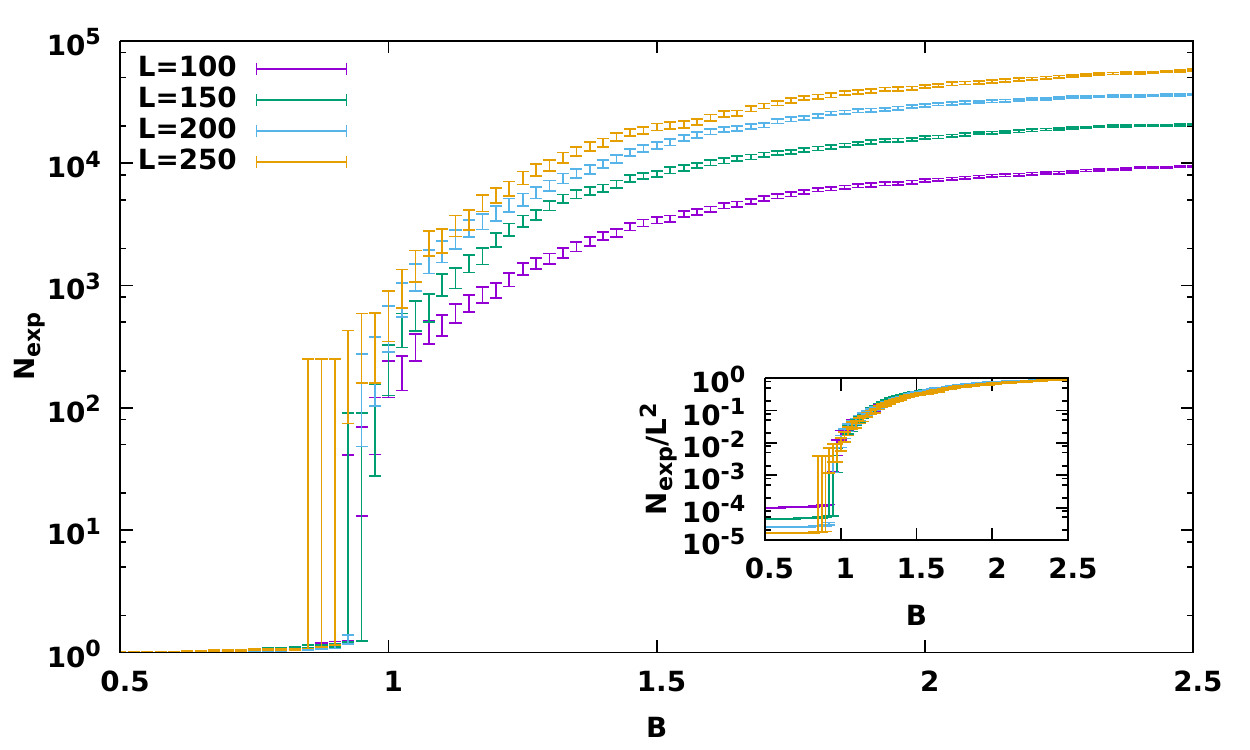}
\end{minipage}
\caption{Two-dimensional deflagration regime. {\em Main plots:} Number of exploded sites $N_{\textrm{exp}}$ as a function of $B$, for dissipation times $\tau=0.7 D/a^2$ (a-c) and $\tau=\infty$ (d-f). From top to bottom, the models with $Q_{i}\propto T_{ci}$ (a,d), $Q_{i}\propto T_{ci}^{2}$ (b,e), and $Q_{i}\propto (T_{ci}+U)^{-1}$ (c,f) are shown. Different data sets correspond to different system sizes $L$. For the $Q\propto T_c$ model, the data are taken from Ref.~\cite{MauroBorisChrisSasha} and are shown here for completeness. {\em Insets:} for short dissipation times (a-c), finite size scaling shows that the transition falls into the universality class of 2D percolation. This is shown by plotting $N_\textrm{exp}L^{-\frac{\gamma}{\nu}}$ as a function of $L^{\frac{1}{\nu}}(B-B_c)/B_c$. For long dissipation times (d-f),  a first-order transition is clearly visible, and the quantity $N_{\textrm{exp}}/L^2$, plotted as a function of $B$, is system size independent after the transition.}
\label{fig:Heat_2D}
\end{figure}

\end{widetext}

\subsubsection{Scaling analysis} 
\label{sub:scaling_analysis}

As our initialization procedure involves randomly sampling an initial site, the final number of exploded sites, $N_{\textrm{exp}}$, corresponds to the \emph{mean cluster size} in the mapping to percolation. 
In order to confirm that the second-order transition lies in the percolative universality class, we perform a finite size scaling of our data, using the following scaling ansatz:
\begin{equation}
\label{eq:scaling}
	N_{\textrm{exp}}(B,L)\propto L^{\frac{\gamma}{\nu}}f\left(L^{\frac{1}{\nu}}\frac{B-B_{c}}{B_{c}}\right),
\end{equation}
where $L$ is the linear size of the sample, and $\gamma$ and $\nu$ are scaling exponents. 
In particular, $\nu$ governs the divergence of the correlation length $\xi$ at the critical point, while $\gamma$ describes the divergence of the mean cluster size. 

To compare our data with percolation theory, we collapse the data at different system sizes for a given disorder model using $B_c$ as a fitting parameter.
For the exponents $\nu$ and $\gamma$ we use the values found in the percolation theory literature: $\nu=\nicefrac{4}{3}$ and $\gamma=\nicefrac{43}{18}$ for $d=2$, and $\nu=0.87\pm0.02$ and $\gamma=1.7\pm0.1$ for $d=3$~\cite{stauffer1994introduction,meester1996continuum}.
This procedure works for $\tau<\tau_c$: in the dissipative regime, the transition belongs to the same universality class as percolation theory. 
For $\tau>\tau_c$, on the other hand, this scaling procedure fails entirely: data at different system sizes simply fall on top of each other for $B<B_c$, and are proportional to $L^d$ for $B>B_c$, with a discontinuous jump for $B=B_c$. The transition is first order.
The tricritical endpoint $\tau_c$ is shown as a red dot in Fig.~\ref{fig:deflagration_phasediagram}.

\begin{figure}[ptb]
\includegraphics[scale=0.7]{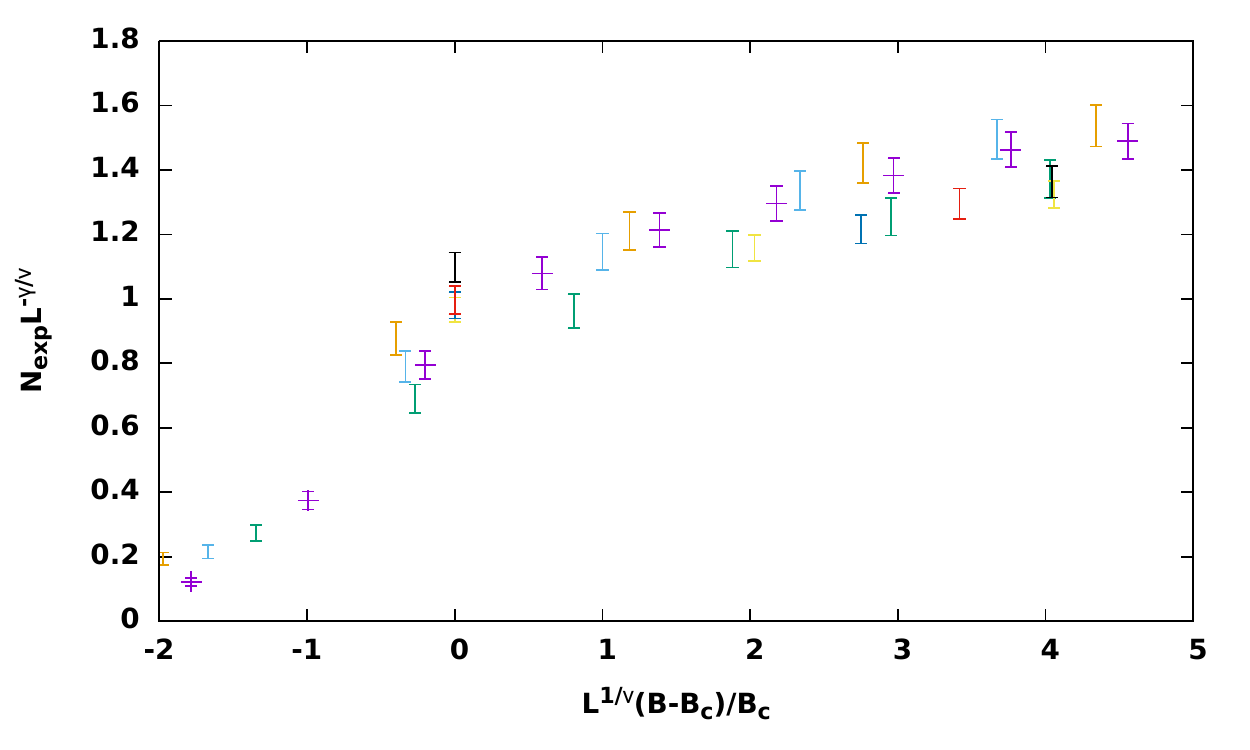}
\includegraphics[scale=0.7]{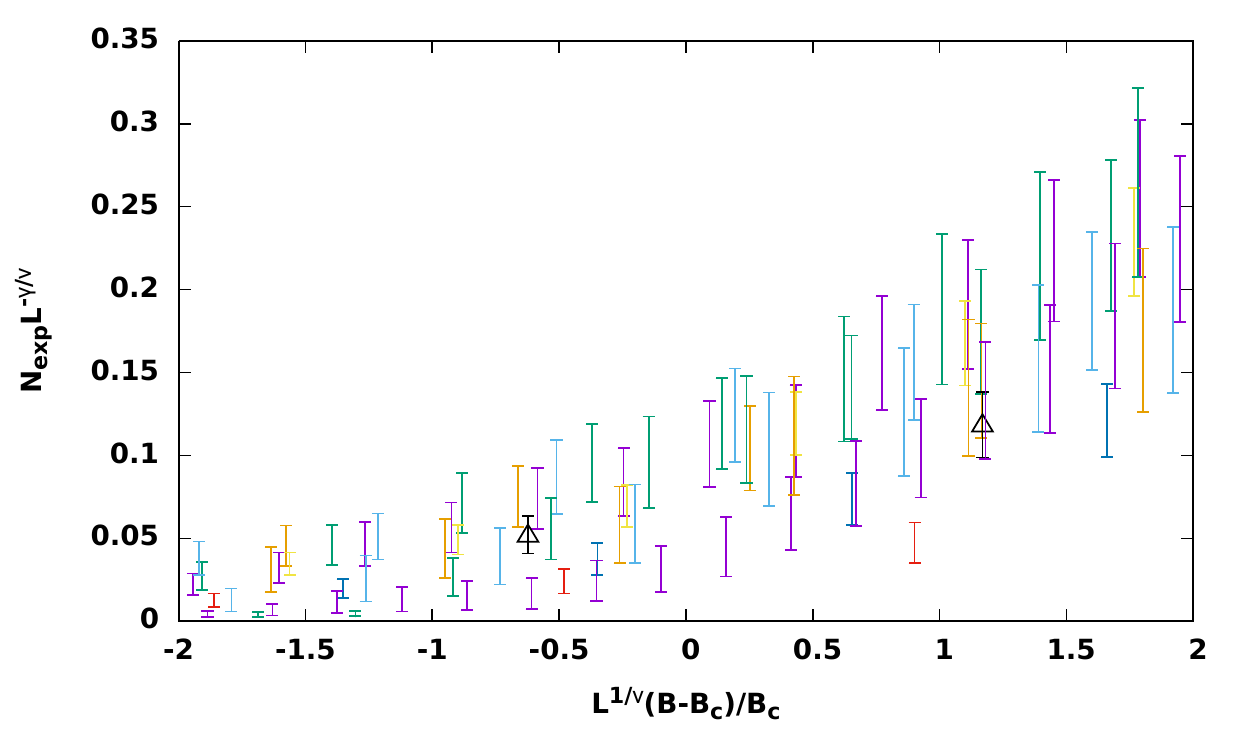}
\caption{Scaling collapse for all models belonging to the percolation theory universality class, in $d=2$ {\em (top)} and $d=3$ {\em (bottom)}. Each color corresponds to a different data set (model and system size). Data from all models follow the same universal function.}
\label{fig:scaling}
\end{figure}

In order to further confirm the robustness of these results, we also collapse data from across all microscopic disorder models. 
This procedure is performed separately for $d=2$ and $d=3$, and the results are shown in Fig.~\ref{fig:scaling}. 
In order to achieve inter-model collapse, two extra fitting parameters, $p$ and $q$, are needed. 
The intermodel scaling ansatz is
\begin{equation}
N_{\textrm{exp}}(B,L)\propto p L^{\frac{\gamma}{\nu}}f\left(qL^{\frac{1}{\nu}}\frac{B-B_{c}}{B_{c}}\right).
\end{equation}
$p$ and $q$ are model dependent, but system size independent, and are of $O(1)$ for all models (ranging between $0.5$ and $4$ in the data shown). 
This is remarkable, since the values of $B_c$ can vary by orders of magnitude as a function of model, $\tau$, and spatial dimension. 
For $d=2$ [Fig.~\ref{fig:scaling}, (a)], the collapse is good, while for $d=3$ [Fig.~\ref{fig:scaling}, (b)] it is still visible, but less remarkable. 
This is most likely because we are restricted to much smaller system sizes in $d=3$, and therefore finite size corrections to scaling are more important.

\subsection{Numerical simulations of the pressure-mediated regime}

For the pressure mediated model, we simulate the wave equation, Eq.~\eqref{eq:wavepropagation}, by discretization in complete analogy to the simulations of the diffusive model.
We choose the time step $\Delta t$ such that
\begin{align}
r\equiv\frac{a\Delta t}{c}=\frac{1}{\sqrt{d}},
\end{align}
satisfies the well-known Courant-Friedrichs-Lewy condition,  which is necessary for the simulation of wave equations to be stable~\cite{NumericalRecipes}.

Explicitly, this is implemented by repeating the following procedure at each time step $\Delta t$, for each site $i$.
\begin{itemize}

\item If site $i$ is still active and $\ensuremath{P({\bf r}_{i},t)\geq P_{c}^{(i)}}$, the site explodes, 
\begin{equation}
P\left(\mathbf{r},t\right)\rightarrow P\left(\mathbf{r},t\right)+Q_i e^{-(\mathbf{r}-\mathbf{r}_i)^2/r_0^2},
\end{equation}
and becomes exhausted.

\item The pressure propagates according to the discretized wave equation,
\begin{align}
	P\left(\mathbf{r}_{i},t+\Delta t\right)&=r^2\sum_{j=1}^{Z}P\left(\mathbf{r}_{j},t\right)+(1-Z r^2)P\left(\mathbf{r}_{i},t\right)\nonumber \\ 
	&-P\left(\mathbf{r}_{i},t-\Delta t\right),
\end{align}
where $Z$ is the coordination number of the lattice, and $j$ labels the nearest neighbors of site $i$. 
\end{itemize}

The simulation is terminated when the pressure waves are too weak to activate any unexploded site.

\begin{widetext}

\begin{figure}
\begin{minipage}[t]{0.4\linewidth}
\includegraphics[scale=0.7]{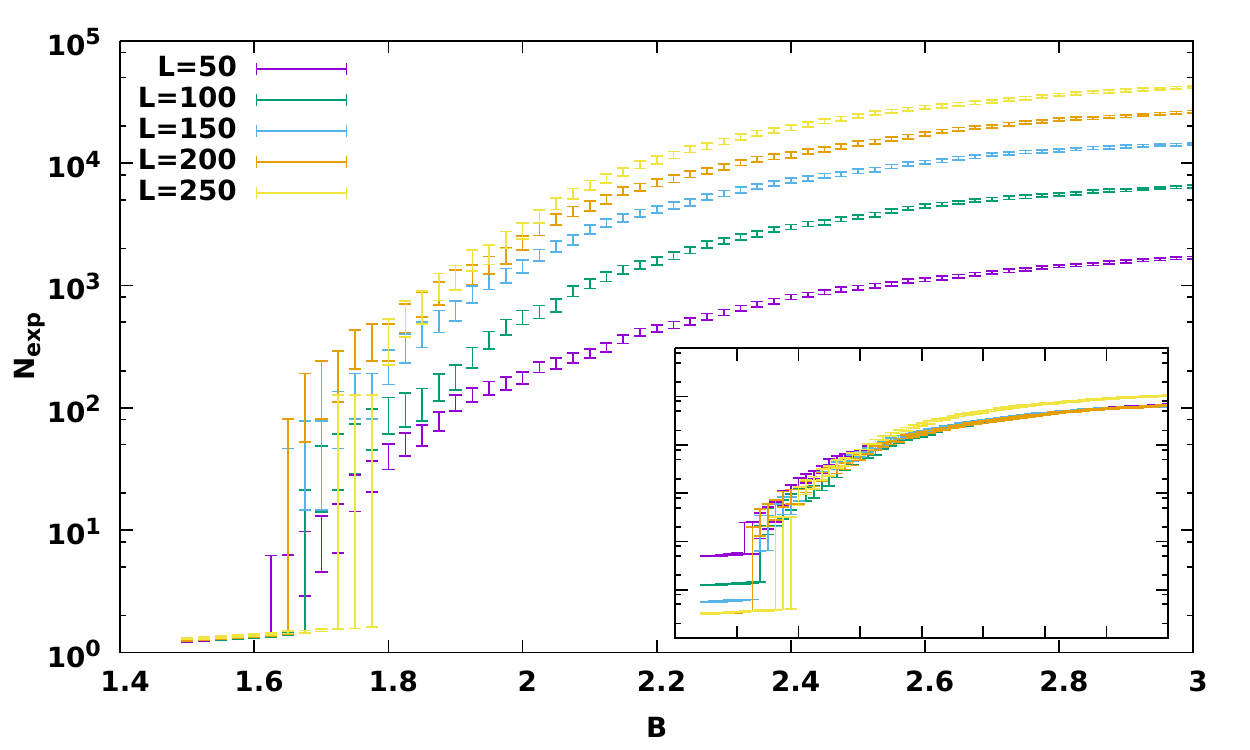}
\includegraphics[scale=0.7]{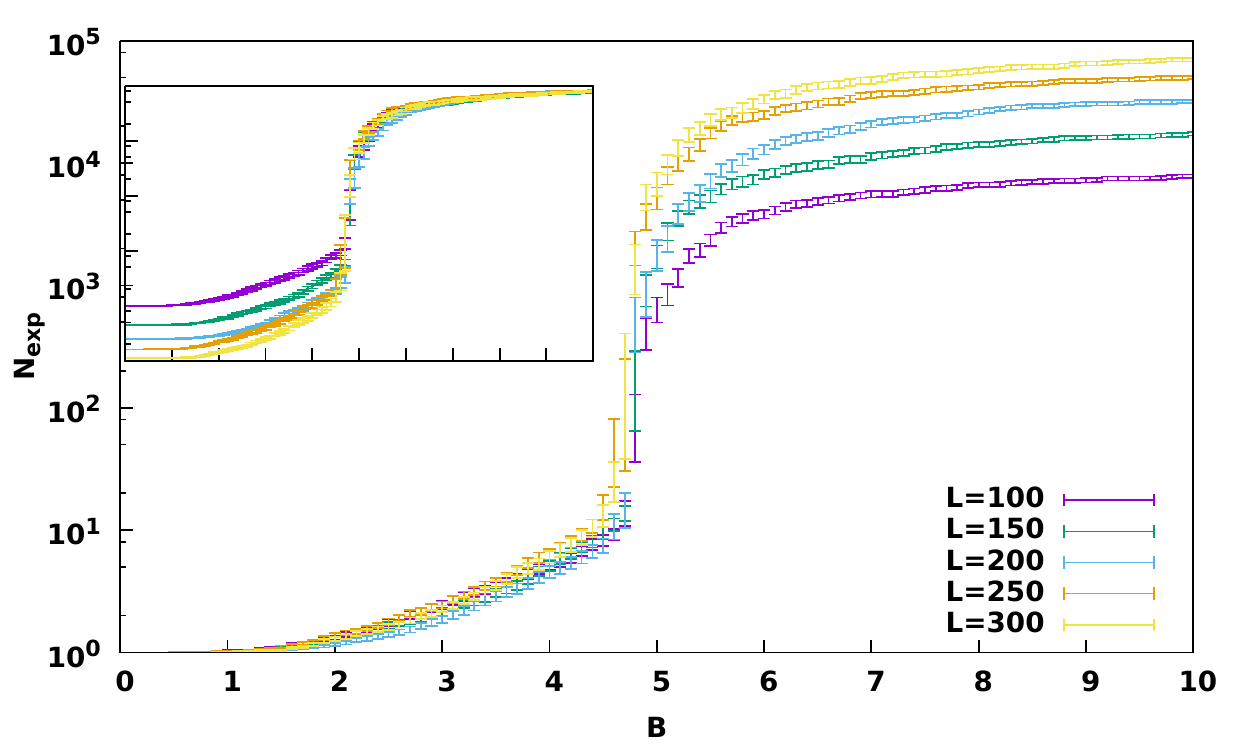}
\includegraphics[scale=0.7]{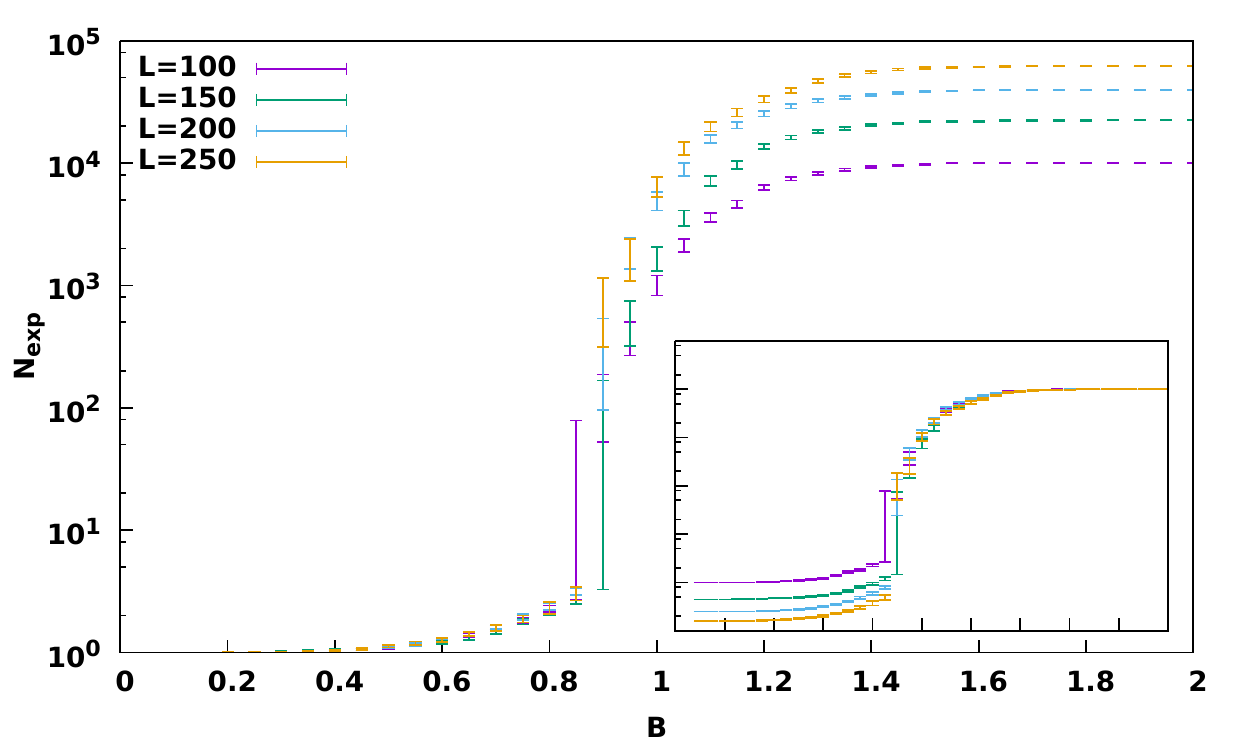}
\end{minipage}
\hspace{1.5cm}
\begin{minipage}[t]{0.4\linewidth}
\includegraphics[scale=0.7]{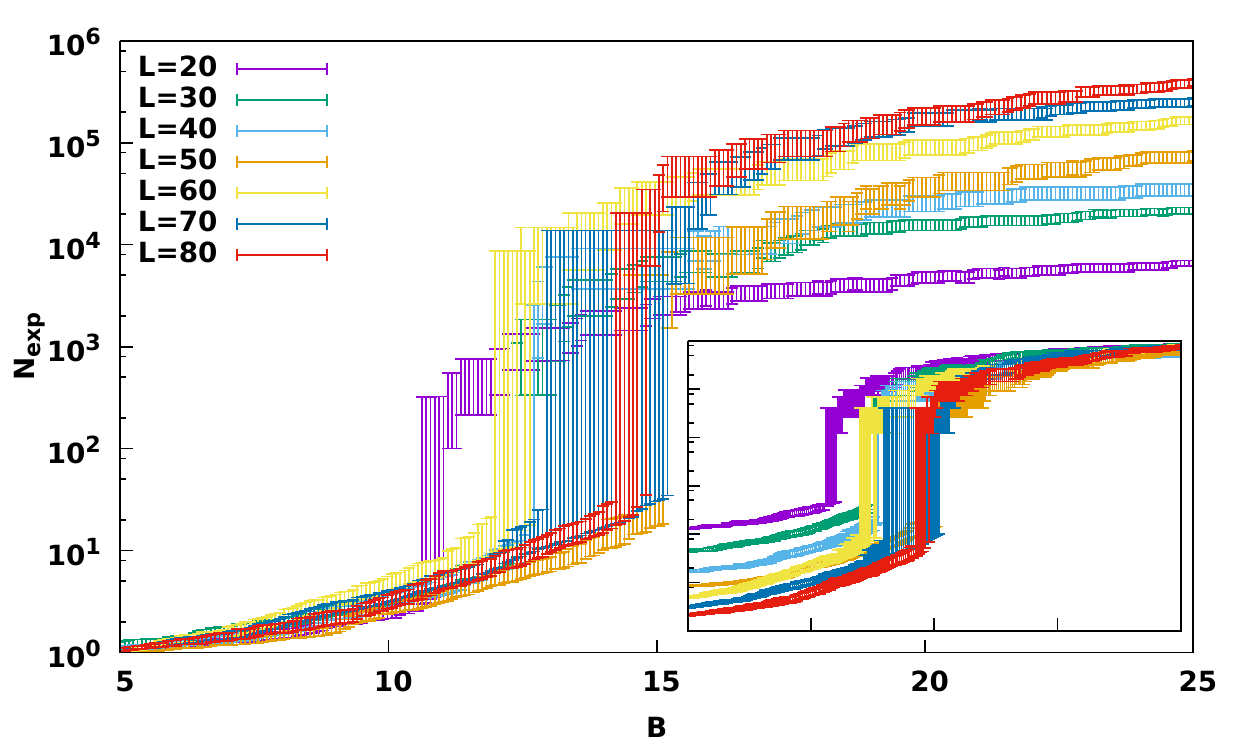}
\includegraphics[scale=0.7]{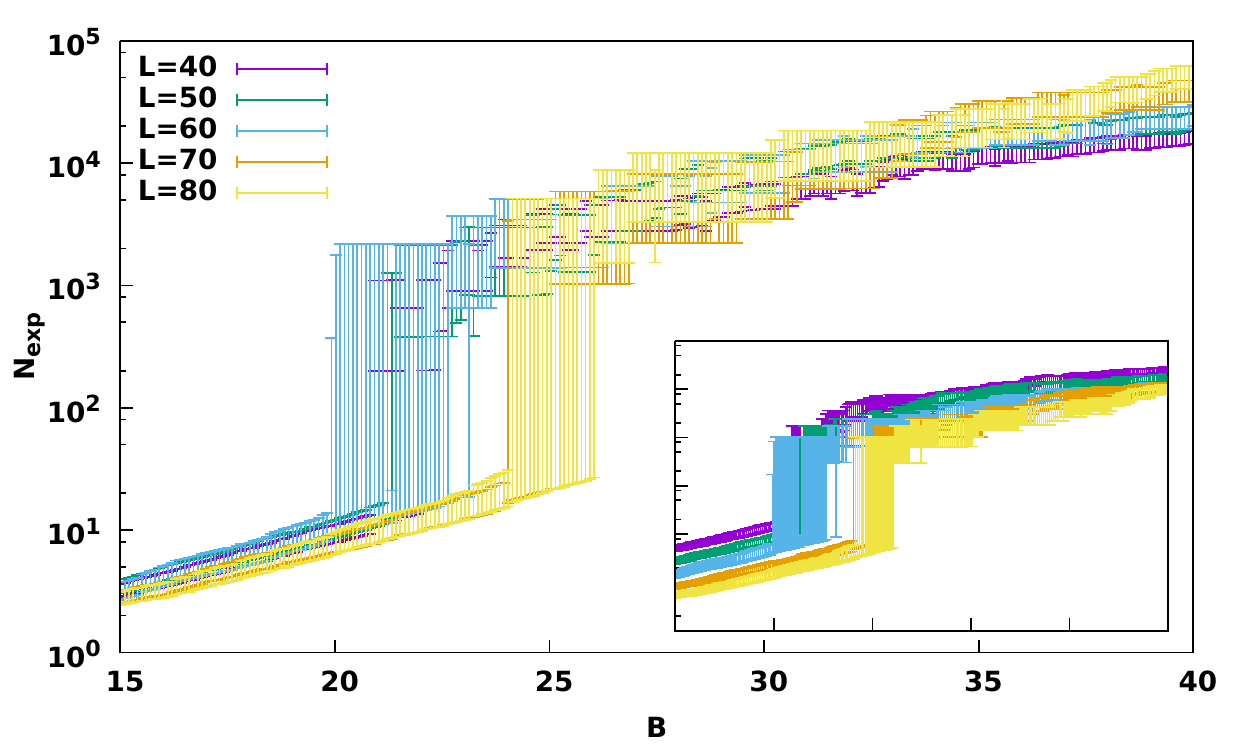}
\includegraphics[scale=0.7]{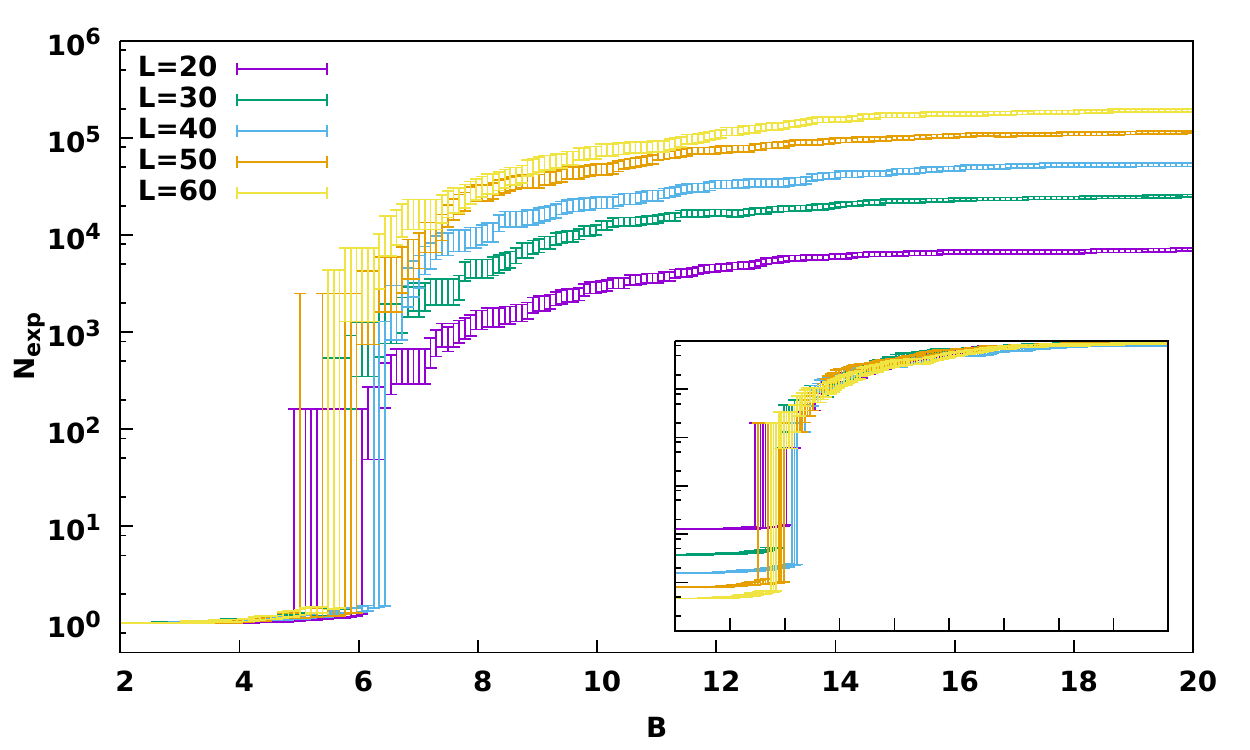}
\end{minipage}
\caption{{\em Main plots:} Sound waves model: number of exploded sites $N_{\textrm{exp}}$ as a function of $B$. Data for both $d=2$ (a-c) and $d=3$ (d-f) are shown. From top to bottom: Model with $Q\propto P_c$ (a,d), $Q\propto P_c^2$ (b,e), and $Q\propto (P_c+U)^{-1}$ (c-f). {\em Insets:} a first order transition is clearly visible. and the quantity $N_{\textrm{exp}}/L^2$, plotted as a function of $B$, is system size independent after the transition}
\label{fig:Soundwaves}
\end{figure}

\end{widetext}

\begin{figure}[ptb]
\includegraphics[scale=0.7]{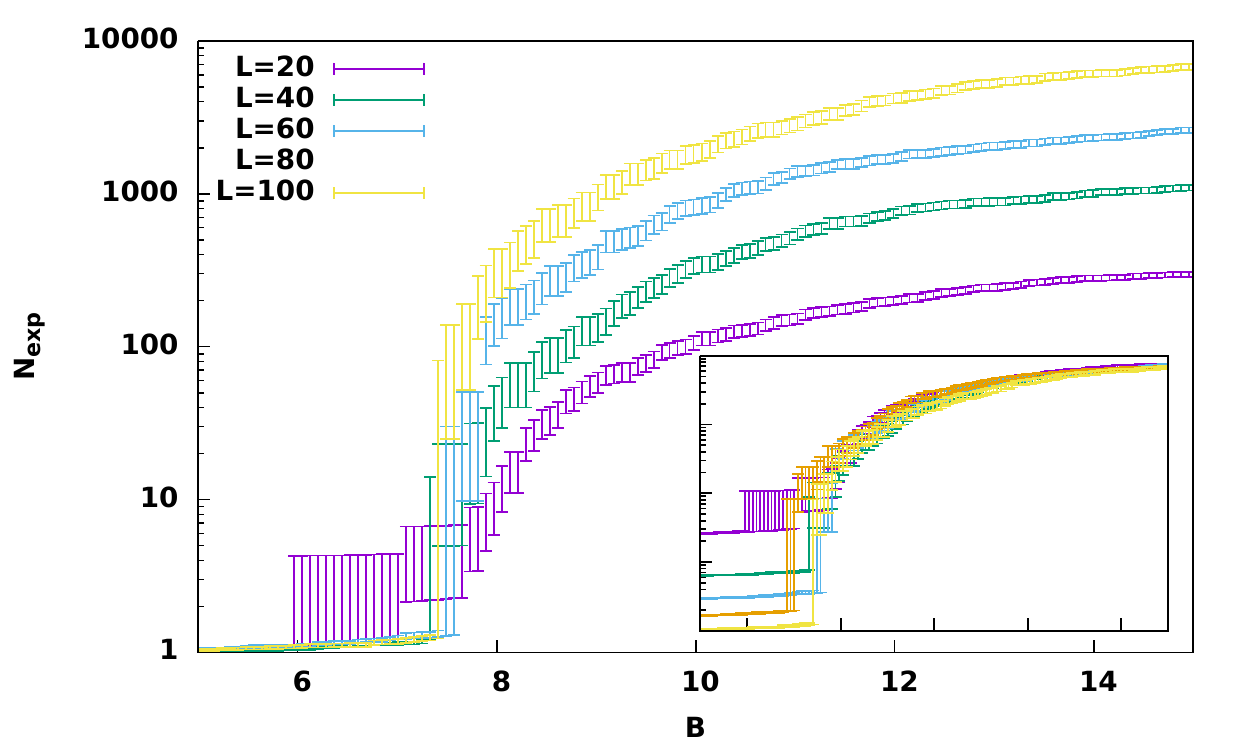}
\includegraphics[scale=0.7]{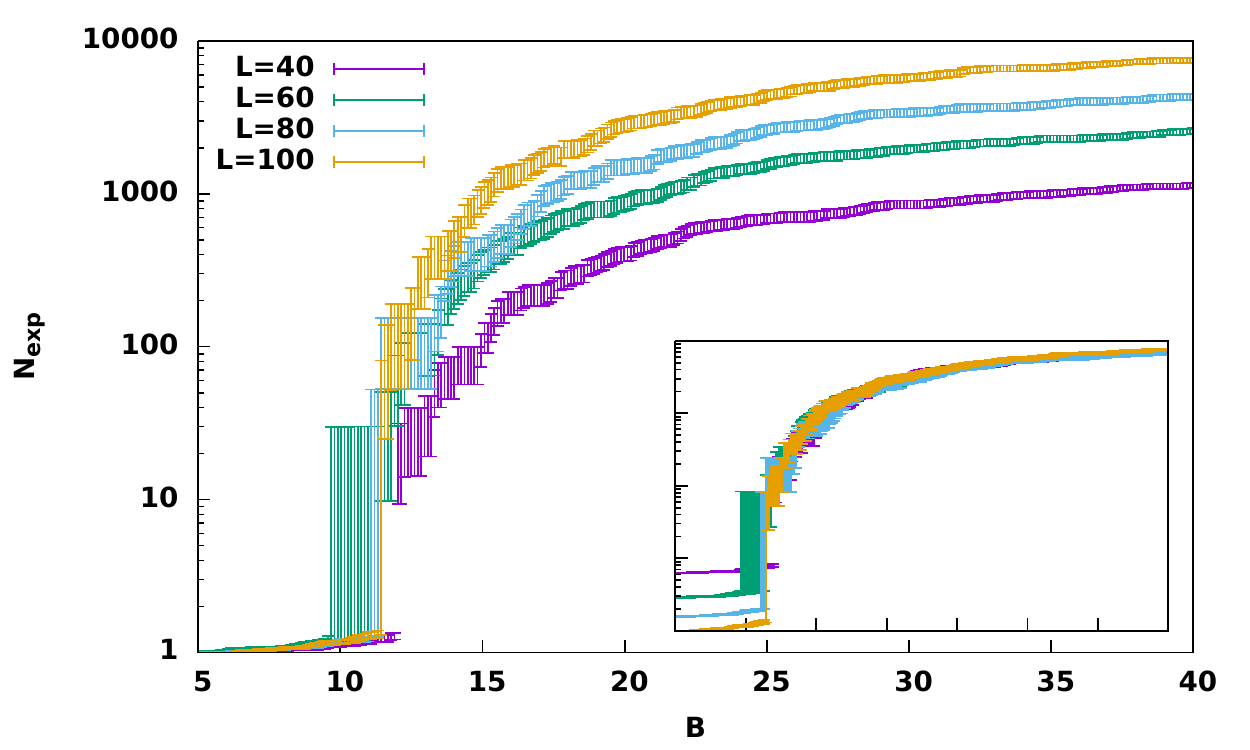}
\includegraphics[scale=0.7]{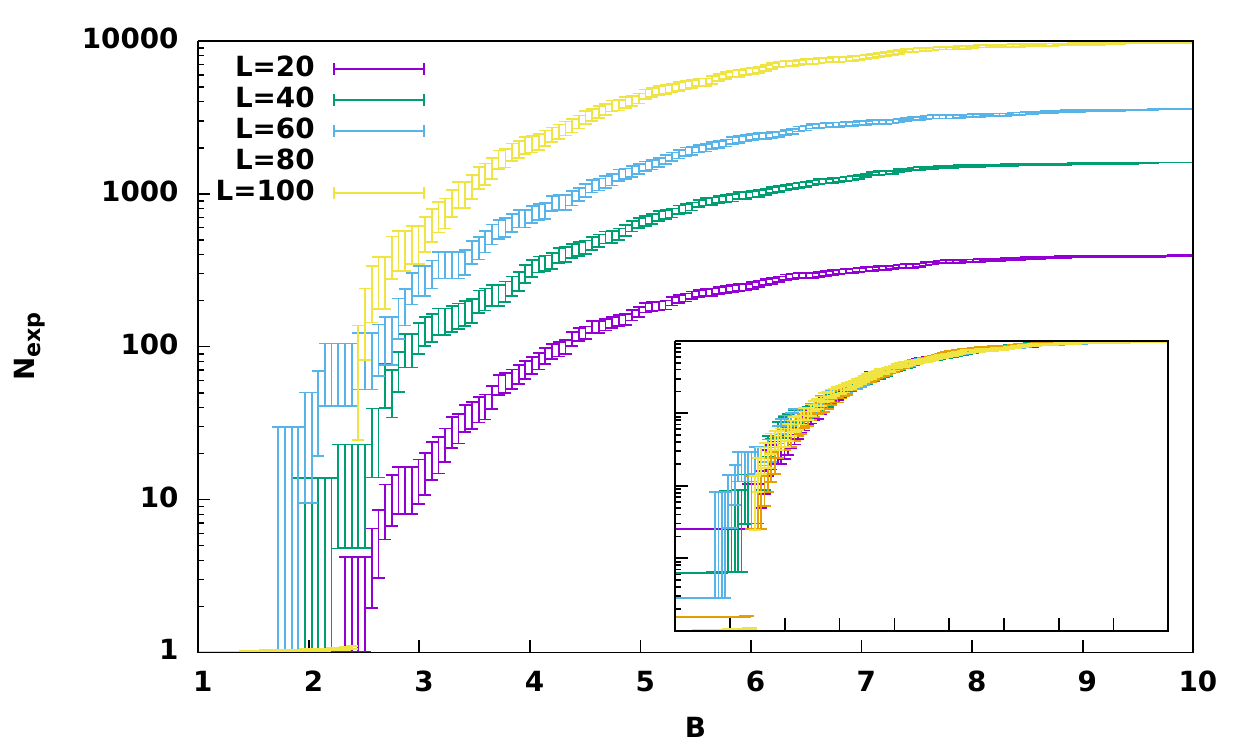}
\caption{{\em Main plots:} Sound waves model, 2D explosive layer embedded in a 3D material, for $Q\propto P_c$ (a), $Q\propto P_c^2$ (b) and $Q\propto(P_c+U)^{-1}$ (c). {\em Insets:} A first order transition is clearly visible, and the quantity $N_{\textrm{exp}}/L^2$, plotted as a function of $B$, is system size independent after the transition}
\label{fig:Sound2Din3D}
\end{figure}

The number of exploded sites $N_\textrm{exp}$ is plotted in Fig.~\ref{fig:Soundwaves} for the two- and three-dimensional sound waves model, and in Fig.~\ref{fig:Sound2Din3D} for the two-dimensional layer embedded in a three-dimensional medium. Since no dissipation is present in the system, the transition is first order in all cases.

\section{Conclusion}
\label{sec:conclusion}

We have considered  disordered explosives containing a system of hot spots, which  can interact either via heat propagation or via sound waves.  We have shown that as a function of  values of parameters  the system exhibits either a first or second order dynamical phase transition, with a tri-critical point dividing the two regimes.  On the qualitative level the phase diagram of the system is universal.  
It is independent of the dimensionality of space, details of the correlation between $T_{ci}$ and $Q_{i}$, and the mechanism of interaction between the hot spots. 
To further corroborate this point we also simulated the detonation of 2D explosive embedded into a 3D host in both the deflagration and the pressure-mediated  regimes. 
The result is again qualitatively consistent with the universal phase diagram.

We believe that the results presented in  Sec.~\ref{sec:combustion_in_the_ballistic_regime}  for the case of hot spots interacting via sound waves should extend as well to the case where the energy is mediated by weak shock waves, as these also propagate ballistically.
In a more general context our results can be applied to systems which exhibit thermal-instability driven avalanches (see, for example,  Refs.~\cite{Rakhmanov,GiaWei}).

One of the limitations of the presented results is the assumption that during the burning process the different parts of the sample do not significantly move apart.  We are planning to consider this aspect of the problem in future work.

We also note that such first-order transition has been reported numerically in the detonation regime as well~\cite{TarverNichols}. However, the authors of Ref.~\cite{TarverNichols} assume the detonation wave to be sufficiently strong to average over all local fluctuations of the system parameters: this can be seen as a mean-field version of the problem addressed in this paper.

{\em Note added in proof - }Additionally, a model similar to ours was used to study the dynamics of front propagation in inhomogeneous systems (see, e.g., Refs.~\cite{Goroshin2011,Lam2017}). These studies are carried in the region of the first-order transition, and do not investigate the nature of the transition itself.

\begin{acknowledgements}
A.B. and B.S. acknowledge financial support from Los Alamos National Laboratory, Contract No. XWA200. 
C.R.L. acknowledges support from the Sloan Foundation through a Sloan Research Fellowship and the NSF through Grant No. PHY-1656234. Any opinion, findings, and conclusions or recommendations expressed in this material are those of the authors and do not necessarily reflect the views of the NSF.
\end{acknowledgements}

\bibliography{explosions}

\begin{thebibliography}{18}
\expandafter\ifx\csname natexlab\endcsname\relax\def\natexlab#1{#1}\fi
\expandafter\ifx\csname bibnamefont\endcsname\relax
  \def\bibnamefont#1{#1}\fi
\expandafter\ifx\csname bibfnamefont\endcsname\relax
  \def\bibfnamefont#1{#1}\fi
\expandafter\ifx\csname citenamefont\endcsname\relax
  \def\citenamefont#1{#1}\fi
\expandafter\ifx\csname url\endcsname\relax
  \def\url#1{\texttt{#1}}\fi
\expandafter\ifx\csname urlprefix\endcsname\relax\def\urlprefix{URL }\fi
\providecommand{\bibinfo}[2]{#2}
\providecommand{\eprint}[2][]{\url{#2}}

\bibitem[{\citenamefont{Landau and Lifshitz}(1987)}]{Landau:1987ab}
\bibinfo{author}{\bibfnamefont{L.~D.} \bibnamefont{Landau}} \bibnamefont{and}
  \bibinfo{author}{\bibfnamefont{E.~M.} \bibnamefont{Lifshitz}},
  \emph{\bibinfo{title}{Fluid Mechanics}} (\bibinfo{publisher}{{Elsevier
  Science \& Technology}}, \bibinfo{year}{1987}), \bibinfo{edition}{2nd} ed.,
  ISBN \bibinfo{isbn}{9780750627672}.

\bibitem[{\citenamefont{Dremin}(1999)}]{dremin1999toward}
\bibinfo{author}{\bibfnamefont{A.}~\bibnamefont{Dremin}},
  \emph{\bibinfo{title}{Toward Detonation Theory}}
  (\bibinfo{publisher}{Springer New York}, \bibinfo{address}{New York, NY},
  \bibinfo{year}{1999}), ISBN \bibinfo{isbn}{1461268192}.

\bibitem[{\citenamefont{Lee}(2008)}]{lee2008the}
\bibinfo{author}{\bibfnamefont{J.}~\bibnamefont{Lee}},
  \emph{\bibinfo{title}{The detonation phenomenon}}
  (\bibinfo{publisher}{Cambridge University Press},
  \bibinfo{address}{Cambridge, UK}, \bibinfo{year}{2008}), ISBN
  \bibinfo{isbn}{0521897238}.

\bibitem[{\citenamefont{Frank-Khamenetski}(1955)}]{Frank-Khamenetski:1955aa}
\bibinfo{author}{\bibfnamefont{D.~A.} \bibnamefont{Frank-Khamenetski}},
  \emph{\bibinfo{title}{Diffusion and Heat Exchange in Chemical Kinetics}}
  (\bibinfo{publisher}{Princeton University Press},
  \bibinfo{address}{Princeton, NJ}, \bibinfo{year}{1955}).

\bibitem[{\citenamefont{Zeldovich}(1980)}]{Zeldovich:1980aa}
\bibinfo{author}{\bibfnamefont{Y.}~\bibnamefont{Zeldovich}},
  \bibinfo{journal}{Combustion and Flame} \textbf{\bibinfo{volume}{39}},
  \bibinfo{pages}{211} (\bibinfo{year}{1980}).

\bibitem[{\citenamefont{Bowden}(1985)}]{bowden1985initiation}
\bibinfo{author}{\bibfnamefont{F.}~\bibnamefont{Bowden}},
  \emph{\bibinfo{title}{Initiation and growth of explosion in liquids and
  solids}} (\bibinfo{publisher}{Cambridge University Press},
  \bibinfo{address}{Cambridge New York}, \bibinfo{year}{1985}), ISBN
  \bibinfo{isbn}{0521312337}.

\bibitem[{\citenamefont{Field}(1992)}]{Field:1992aa}
\bibinfo{author}{\bibfnamefont{J.~E.} \bibnamefont{Field}},
  \bibinfo{journal}{Accounts of Chemical Research}
  \textbf{\bibinfo{volume}{25}}, \bibinfo{pages}{489} (\bibinfo{year}{1992}),
  \urlprefix\url{http://dx.doi.org/10.1021/ar00023a002}.

\bibitem[{\citenamefont{Tarver et~al.}(1996)\citenamefont{Tarver, Chidester,
  and Nichols~III}}]{Tarver:1996aa}
\bibinfo{author}{\bibfnamefont{C.~M.} \bibnamefont{Tarver}},
  \bibinfo{author}{\bibfnamefont{S.~K.} \bibnamefont{Chidester}},
  \bibnamefont{and} \bibinfo{author}{\bibfnamefont{A.~L.}
  \bibnamefont{Nichols~III}}, \bibinfo{journal}{The Journal of Physical
  Chemistry} \textbf{\bibinfo{volume}{100}}, \bibinfo{pages}{5794}
  (\bibinfo{year}{1996}), \urlprefix\url{http://dx.doi.org/10.1021/jp953123s}.

\bibitem[{\citenamefont{McGrane et~al.}(2009)\citenamefont{McGrane, Grieco,
  Ramos, Hooks, and Moore}}]{McGrane:2009aa}
\bibinfo{author}{\bibfnamefont{S.~D.} \bibnamefont{McGrane}},
  \bibinfo{author}{\bibfnamefont{A.}~\bibnamefont{Grieco}},
  \bibinfo{author}{\bibfnamefont{K.~J.} \bibnamefont{Ramos}},
  \bibinfo{author}{\bibfnamefont{D.~E.} \bibnamefont{Hooks}}, \bibnamefont{and}
  \bibinfo{author}{\bibfnamefont{D.~S.} \bibnamefont{Moore}},
  \bibinfo{journal}{J. Appl. Phys.} \textbf{\bibinfo{volume}{105}},
  \bibinfo{pages}{073505} (\bibinfo{year}{2009}).

\bibitem[{\citenamefont{Schiulaz et~al.}(2017)\citenamefont{Schiulaz, Laumann,
  Balatsky, and Spivak}}]{MauroBorisChrisSasha}
\bibinfo{author}{\bibfnamefont{M.}~\bibnamefont{Schiulaz}},
  \bibinfo{author}{\bibfnamefont{C.~R.} \bibnamefont{Laumann}},
  \bibinfo{author}{\bibfnamefont{A.~V.} \bibnamefont{Balatsky}},
  \bibnamefont{and} \bibinfo{author}{\bibfnamefont{B.~Z.}
  \bibnamefont{Spivak}}, \bibinfo{journal}{Phys. Rev. E}
  \textbf{\bibinfo{volume}{95}}, \bibinfo{pages}{032103}
  (\bibinfo{year}{2017}),
  \urlprefix\url{https://link.aps.org/doi/10.1103/PhysRevE.95.032103}.

\bibitem[{\citenamefont{Stauffer}(1994)}]{stauffer1994introduction}
\bibinfo{author}{\bibfnamefont{D.}~\bibnamefont{Stauffer}},
  \emph{\bibinfo{title}{Introduction to percolation theory}}
  (\bibinfo{publisher}{Taylor \& Francis}, \bibinfo{address}{London Bristol,
  PA}, \bibinfo{year}{1994}), ISBN \bibinfo{isbn}{0748402535}.

\bibitem[{\citenamefont{Meester}(1996)}]{meester1996continuum}
\bibinfo{author}{\bibfnamefont{R.}~\bibnamefont{Meester}},
  \emph{\bibinfo{title}{Continuum percolation}} (\bibinfo{publisher}{Cambridge
  University Press}, \bibinfo{address}{Cambridge New York},
  \bibinfo{year}{1996}), ISBN \bibinfo{isbn}{052147504X}.

\bibitem[{\citenamefont{Press et~al.}(2007)\citenamefont{Press, Teukolsky,
  Vetterling, and Flannery}}]{NumericalRecipes}
\bibinfo{author}{\bibfnamefont{W.~H.} \bibnamefont{Press}},
  \bibinfo{author}{\bibfnamefont{S.~A.} \bibnamefont{Teukolsky}},
  \bibinfo{author}{\bibfnamefont{W.~T.} \bibnamefont{Vetterling}},
  \bibnamefont{and} \bibinfo{author}{\bibfnamefont{B.~P.}
  \bibnamefont{Flannery}}, \emph{\bibinfo{title}{Numerical Recipes}}
  (\bibinfo{publisher}{Cambridge University Press}, \bibinfo{year}{2007}).

\bibitem[{\citenamefont{Rakhmanov et~al.}(2004)\citenamefont{Rakhmanov,
  Shantsev, Galperin, and Johansen}}]{Rakhmanov}
\bibinfo{author}{\bibfnamefont{A.~L.} \bibnamefont{Rakhmanov}},
  \bibinfo{author}{\bibfnamefont{D.~V.} \bibnamefont{Shantsev}},
  \bibinfo{author}{\bibfnamefont{Y.~M.} \bibnamefont{Galperin}},
  \bibnamefont{and} \bibinfo{author}{\bibfnamefont{T.~H.}
  \bibnamefont{Johansen}}, \bibinfo{journal}{Phys. Rev. B}
  \textbf{\bibinfo{volume}{70}}, \bibinfo{pages}{224502}
  (\bibinfo{year}{2004}),
  \urlprefix\url{https://link.aps.org/doi/10.1103/PhysRevB.70.224502}.

\bibitem[{\citenamefont{Chern}(2017)}]{GiaWei}
\bibinfo{author}{\bibfnamefont{G.-W.} \bibnamefont{Chern}},
  \bibinfo{journal}{Journal of Physics: Condensed Matter}
  \textbf{\bibinfo{volume}{29}}, \bibinfo{pages}{044004}
  (\bibinfo{year}{2017}),
  \urlprefix\url{http://stacks.iop.org/0953-8984/29/i=4/a=044004}.

\bibitem[{\citenamefont{Nichols~III and Tarver}(2002)}]{TarverNichols}
\bibinfo{author}{\bibfnamefont{A.~L.} \bibnamefont{Nichols~III}}
  \bibnamefont{and} \bibinfo{author}{\bibfnamefont{C.~M.}
  \bibnamefont{Tarver}}, \bibinfo{journal}{Proceedings of the 12th
  International Detonation Symposium, San Diego, CA}  (\bibinfo{year}{2002}).

\bibitem[{\citenamefont{Goroshin et~al.}(2011)\citenamefont{Goroshin, Tang, and
  Higgins}}]{Goroshin2011}
\bibinfo{author}{\bibfnamefont{S.}~\bibnamefont{Goroshin}},
  \bibinfo{author}{\bibfnamefont{F.-D.} \bibnamefont{Tang}}, \bibnamefont{and}
  \bibinfo{author}{\bibfnamefont{A.~J.} \bibnamefont{Higgins}},
  \bibinfo{journal}{Phys. Rev. E} \textbf{\bibinfo{volume}{84}},
  \bibinfo{pages}{027301} (\bibinfo{year}{2011}),
  \urlprefix\url{https://link.aps.org/doi/10.1103/PhysRevE.84.027301}.

\bibitem[{\citenamefont{Lam et~al.}(2017)\citenamefont{Lam, Mi, and
  Higgins}}]{Lam2017}
\bibinfo{author}{\bibfnamefont{F.}~\bibnamefont{Lam}},
  \bibinfo{author}{\bibfnamefont{X.}~\bibnamefont{Mi}}, \bibnamefont{and}
  \bibinfo{author}{\bibfnamefont{A.~J.} \bibnamefont{Higgins}},
  \bibinfo{journal}{Phys. Rev. E} \textbf{\bibinfo{volume}{96}},
  \bibinfo{pages}{013107} (\bibinfo{year}{2017}),
  \urlprefix\url{https://link.aps.org/doi/10.1103/PhysRevE.96.013107}.

\end{thebibliography}

\end{document}